\documentclass[12pt]{article}
\usepackage[OT1]{fontenc}
\usepackage[utf8]{inputenc}
\usepackage{tocloft}
\usepackage{amsfonts}
\usepackage{amsmath}
\usepackage{amssymb}
\usepackage{array}
\usepackage{slashed}
\usepackage{bigints}
\usepackage{pifont}
\usepackage{bm}
\usepackage{bbm}
\usepackage{upgreek}
\usepackage{booktabs}
\usepackage[nosort]{cite}
\usepackage[table,dvipsnames]{xcolor}
\usepackage{dsfont}
\usepackage{float}
\usepackage{framed}
\usepackage{graphicx}
\usepackage{indentfirst}
\usepackage{mathrsfs}
\usepackage{multirow}
\usepackage{pdflscape}
\usepackage{setspace}
\usepackage{subdepth}
\usepackage{subfig}
\usepackage{titlesec}
\usepackage{wrapfig}
\usepackage[all]{xy}
\usepackage{young}
\usepackage[vcentermath]{youngtab}
\usepackage{relsize}
\usepackage{stackengine}
\usepackage{verbatim}
\usepackage{slashed}
\usepackage{tikz}
\usetikzlibrary{arrows,decorations.markings,cd}

\usepackage{hyperref}
\hypersetup{colorlinks=true}
\hypersetup{linkcolor=black}
\hypersetup{citecolor=black}
\hypersetup{urlcolor=black}

\numberwithin{equation}{section}

\usepackage[left=2.5cm,right=2.5cm,top=2.5cm,bottom=3cm]{geometry}
\fontdimen2\font=1.2\fontdimen2{\jot}{5pt}

\newlength{\bibitemsep}
\newlength{\bibparskip}
\let\oldthebibliography\thebibliography
\renewcommand\thebibliography[1]{%
  \oldthebibliography{#1}%
  \setlength{\parskip}{\bibitemsep}%
  \setlength{\itemsep}{\bibparskip}%
}

\titleformat{\section}{\bfseries}{\thesection.}{4pt}{}
\titlespacing{\section}{0pt}{20pt}{6pt}

\titleformat{\subsection}{\normalfont\itshape}{\thesubsection.}{4pt}{}
\titlespacing{\subsection}{0pt}{15pt}{6pt}

\titleformat{\subsubsection}{\normalfont\itshape}{\thesubsubsection.}{4pt}{}
\titlespacing{\subsubsection}{0pt}{15pt}{6pt}

\titleformat{\paragraph}{\normalfont\itshape}{\theparagraph.}{4pt}{}
\titlespacing{\paragraph}{0pt}{15pt}{6pt}

\renewcommand{\tilde}{\widetilde}

\renewcommand{\hat}{\widehat}

\renewcommand{\bar}{\overline}

\newcommand{\half}{\frac{1}{2}}

\DeclareMathOperator{\Tr}{Tr}

\DeclareMathAlphabet{\mathbfsf}{OT1}{cmss}{bx}{n}

\newcommand{\Z}{{\mathbb Z}}

\newcommand{\bR}{\mathbb{R}}
\newcommand{\bZ}{\mathbb{Z}}

\newcommand{\cA}{\mathcal{A}}

\newcommand{\cD}{\mathcal{D}}

\newcommand{\cI}{\mathcal{I}}

\newcommand{\cM}{\mathcal M}

\newcommand{\cY}{\mathcal{Y}}

\newcommand{\ed}{\,.}
\newcommand{\ec}{\,,}

\newcommand{\be}{\begin{equation}}
\newcommand{\ee}{\end{equation}}
\newcommand{\beq}{\begin{equation}}
\newcommand{\eeq}{\end{equation}}

\newcommand{\zero}{^{(0)}}

\newcommand{\one}{^{(1)}}
\newcommand{\two}{^{(2)}}

\allowdisplaybreaks 

\DeclareFontShape{OT1}{cmr}{mx}{n}%
{<->cmr10}{}
\newcommand{\mytitlefont}{\fontseries{mx}\selectfont}
\DeclareMathAlphabet{\titlemath}{OT1}{cmr}{mx}{n}

\begin{document}

%
\begin{titlepage}
\begin{center}
~\\[1.5cm]
{\fontsize{27pt}{0pt} \mytitlefont Continuous Generalized Symmetries \vskip .3cm in Three Dimensions}
~\\[1.25cm]
Jeremias Aguilera Damia, Riccardo Argurio, and Luigi Tizzano
~\\[0.5cm]
{~{\it Physique Th\'eorique et Math\'ematique and International Solvay Institutes\\
Universit\'e Libre de Bruxelles; C.P. 231, 1050 Brussels, Belgium}}

~\\[1.25cm]
			
\end{center}
\noindent We present a class of three-dimensional quantum field theories 
whose ordinary global symmetries mix with higher-form symmetries to form a continuous $2$-group. All these models can be obtained by performing a gauging procedure in a parent theory revealing a 't Hooft anomaly in the space of coupling constants when suitable compact scalar background fields are activated. Furthermore, the gauging procedure also implies that our main example has infinitely many non-invertible global symmetries. These can be obtained by dressing the continuous symmetry operators with topological quantum field theories. Finally, we comment on the holographic realization of both $2$-group global symmetries and non-invertible symmetries discussed here by introducing a corresponding four-dimensional bulk description in terms of dynamical gauge fields.

\vfill 
\begin{flushleft}
June 2022
\end{flushleft}
\end{titlepage}
%
		
	
\setcounter{tocdepth}{3}
\renewcommand{\cfttoctitlefont}{\large\bfseries}
\renewcommand{\cftsecaftersnum}{.}
\renewcommand{\cftsubsecaftersnum}{.}
\renewcommand{\cftsubsubsecaftersnum}{.}
\renewcommand{\cftdotsep}{6}
\renewcommand\contentsname{\centerline{Contents}}
	
\tableofcontents
	
\section{Introduction}\label{intro}
According to \cite{GKSW} global symmetries in quantum field theory are realized by topological operators. In this context an ordinary $0$-form global symmetry $G\zero$  is thus associated to a topological operator (equivalently a symmetry defect) of codimension-$(0+1)$ while a $p$-form global symmetry $G^{(p)}$ is associated with a topological operator of codimension-$(p+1)$.\footnote{We denote the $p$-form degree of an object by a superscript $^{(p)}$.}

In this work we will be primarily interested in quantum field theories with a collection of continuous $0$-form symmetries $G\zero$ and $1$-form symmetries $G\one$. In all these theories the topological nature of the corresponding symmetry defects encodes current conservation. Many theories in this class exhibit a new surprising feature. Namely, the conserved $2$-form currents $J\two$ appear in the operator product expansion of two $1$-form currents $j\one$.  This implies that the conservation of $J\two$ and $j\one$ at separated points does not necessarily imply that the symmetry groups $G\one$ and $G\zero$ act independently on the theory but rather they mix in a more general global symmetry structure which is known as a ``$2$-group global symmetry" \cite{Kapustin:2013uxa, Sharpe:2015mja, Thorngren:2015gtw, Tachikawa:2017gyf,CDI,Delcamp:2018wlb,Benini:2018reh}.\footnote{See \cite{BaezLauda1} for a mathematical definition of a $2$-group structure.}

To understand how $2$-group global symmetries can appear in quantum field theory it is instructive to consider a basic model, analyzed in \cite{CDI}, consisting of $2N_f$ massless Weyl fermions in four dimensions denoted by $(\psi^i, \tilde{\psi}^{\,\tilde{i}})$ with $i,\tilde{i}=1,\cdots, N_f$. Under the global symmetries $SU(N_f)\zero_L\times SU(N_f)\zero_R \times U(1)\zero_C$, $\psi$ transforms in the fundamental of $SU(N_f)\zero_L$ and it is neutral under $SU(N_f)\zero_R$ while $\tilde{\psi}$ transforms in the fundamental of $SU(N_f)\zero_R$ and it is neutral under $SU(N_f)\zero_L$. For simplicity, we assume that the charges of $\psi$ and $\tilde{\psi}$ under $U(1)\zero_C$ are (respectively) $+1$ and $-1$. 

All perturbative 't Hooft anomalies for this theory are completely characterized in terms of a six-form anomaly polynomial $\cI^{(6)}$ satisfying the following descent relations
\be
\delta {\cal I}^{(6)}= d{\cal I}^{(6)}=0\ec \quad {\cal I}^{(6)}= d{\cal I}^{(5)}\ec \quad \delta{\cal I}^{(5)}= d{\cal A}^{(4)}\ec
\label{eq:descent1}
\ee   
where $\cA^{(4)}$ is a functional that measures the anomalies.\footnote{The second step in \eqref{eq:descent1} is notoriously ambiguous. Namely we can always shift ${\cal I}^{(5)}$ by an exact $5$-form. This descends to the  freedom of adding local counterterms which may affect the presentation of the anomaly ${\cal A}^{(4)}$.} Our main focus lies in the mixed anomalies sector contained in $\cI^{(6)}$ namely
\be\label{I6mixed}
\cI_{\textrm{Mixed}}^{(6)}= \frac{1}{ 4\pi^2}\left(-\frac{k_{L^2C}}{2!} \Tr(F_L^{(2)}\wedge F_L^{(2)})\wedge F_C^{(2)} -\frac{k_{R^2C}}{2!} \Tr(F_R^{(2)}\wedge F_R^{(2)})\wedge F_C^{(2)}\right)\ec
\ee
where $F\two_{L,R}=dA\one_{L,R}+A_{L,R}\one\wedge A_{L,R}\one$ and $F\two_C=dA\one_C$ are all background field strengths for the associated global symmetry. Looking at \eqref{I6mixed}, the anomalous shifts of the partition function under background gauge variations is given by 
\be\label{notsmart}
Z[A\one_{L}\to A\one_{L}+d\lambda\zero_{L},A\one_R,C\one] = Z[A\one_{L},A\one_R,C\one]\exp\left(-\frac{ik_{L^2C}}{8\pi^2}\int\Tr\left(\lambda\zero_{L}dA_{L}\one\right)\wedge F\two_C\right)\ec
\ee
and similarly for $A\one_{R}\to A\one_{R}+d\lambda\zero_{R}$. Note that the $U(1)\zero_C$ charges of the massless Weyl fermions guarantee that the theory does not suffer from a $U(1)\zero_C$ cubic anomaly i.e. $k_{C^3}=0$.
This implies that we can safely gauge the global symmetry $U(1)\zero_C$ by promoting $C^{(1)}\to c^{(1)}$ ($F_C^{(2)}\to f^{(2)}_c$) with $c^{(1)}$ now being a \emph{dynamical} field. The result of this gauging procedure is that we obtain a new theory known as four-dimensional massless quantum electrodynamics (QED). However, gauging $U(1)\zero_C$ also leads to a confusing effect. The phase on the right hand side of \eqref{notsmart} is \emph{not} a 't Hooft anomaly, at the same time it also \emph{does not} violate the $SU(N_f)\zero_{L, R}$ symmetries. Hence, it must be compensated within the path integral by postulating new background transformation rules for the symmetries at our disposal. 

The crucial point is that gauging a $U(1)\zero_C$ symmetry in four dimensions introduces a new $1$-form global symmetry $U(1)_B\one$ whose conserved current is
\be
J\two_B = \frac{1}{2\pi}*f\two_c\ec
\ee
that can be coupled to a background $2$-form gauge field $B^{(2)}$ as follows
\be
\int B^{(2)}\wedge * J_B^{(2)} =\frac{1}{2\pi}\int B^{(2)}\wedge f_c^{(2)}\ .
\label{eq:4dmag1}
\ee 
To eliminate the problematic transformation appearing in \eqref{notsmart} we first define the coupled system
\begin{align}
    \tilde Z[A_{L,R}^{(1)},B^{(2)}] \equiv \int {\cal D}c^{(1)}\ Z[A_{L,R}^{(1)},c^{(1)}]\ e^{\frac{i}{2\pi}\int B^{(2)}\wedge f_c^{(2)}}\ec
\end{align}
and then postulate the following background gauge transformations\footnote{In analogy with conventional background gauge fields, a $B\two$ $2$-form gauge field subject to \eqref{eq:2grtrans} can be
thought of as a $2$-connection on an associated $2$-bundle \cite{Baez:2004in,SchreiberWaldorf}.}
\be
A_{L,R}^{(1)}\to A_{L,R}^{(1)}+d\lambda_{L,R}^{(0)}\ec \quad B^{(2)}\to B^{(2)}+d\Lambda^{(1)}_B+\frac{k_{L^2C,R^2C}}{4\pi} \Tr\left(\lambda^{(0)}_{L,R} dA_{L,R}^{(1)}\right)\ed
\label{eq:2grtrans}
\ee 
Under these new rules the problematic phase appearing in \eqref{notsmart} is no longer an issue and the partition function is completely gauge invariant
\be
\delta \tilde Z[A_{L,R}\one,B\two] = 0\ec
\ee
We conclude that massless QED coupled to the background fields \eqref{eq:2grtrans} enjoys a $2$-group global symmetry denoted by
\be
\left(SU(N_f)\zero_L \times SU(N_f)\zero_R\right) \times_{\kappa_L = k_{L^2C}, \kappa_R=k_{R^2C}} \times U(1)\one_B\ec
\ee
where $\kappa_L, \kappa_R \in \bZ$ are known as $2$-group structure constants. In the last few years it has been understood that $2$-group global symmetries are commonplace in quantum field theory. See \cite{ Hsin:2020nts,Gukov:2020btk,Cordova:2020tij,DelZotto:2020sop,Apruzzi:2021vcu,Lee:2021crt,Apruzzi:2021mlh} for a large set of examples in $d\geq 4$.

In this paper, we are interested in generalizing the above gauging procedure to describe new examples of continuous $2$-group global symmetries in three-dimensional quantum field theories.
However our program faces two immediate obstacles. First, in three dimensions continuous zero-form global symmetries do not admit 't Hooft anomalies of the same type discussed in the QED example.\footnote{A $U(1)\zero$ global symmetry in three dimensions may have a mixed 't Hooft anomaly with a $U(1)\one$ symmetry as described in  \cite{Delacretaz:2019brr}. These sort of mixed anomalies are interesting due to their close relation to spontaneous symmetry breaking patterns. However, they do not lead to $2$-group global symmetries.} Second, gauging a $U(1)\zero_C$ symmetry is no longer associated to a dual magnetic $1$-form symmetry $U(1)\one_B$. In fact, gauging a $U(1)^{(p)}$ symmetry gives rise to a dual $\tilde U(1)^{(d-3-p)}$ symmetry. Crucially, the resolution for both of these problems relies on a recent generalization of the notion of 't Hooft anomaly that has been put forward in \cite{Seiberg:2018ntt,Cordova:2019jnf,Cordova:2019uob} (see also \cite{Hsin:2020cgg,Kapustin:2020eby,Choi:2022odr} for similar constructions which have been referred to as higher Berry phases).\footnote{Continuous $G\zero$ symmetries in three dimensions can have 't Hooft anomalies whose coefficients take a finite set of integer values. A detailed study of these anomalies can be found in \cite{Benini:2017dus}. (See also \cite{Genolini:2022mpi, Bhardwaj:2022dyt} for some recent applications to $3d$ supersymmetric theories). It can be shown that gauging a discrete subgroup of a symmetry participating in such an anomaly gives rise to \emph{discrete} $2$-group global symmetries. Since our focus is on theories with continuous magnetic $1$-form symmetries we will not pursue this idea further in this paper.}  

The traditional anomaly paradigm can in fact be extended to analyze how the partition function of a given quantum field theory can depend on background scalar fields varying over spacetime. These define a ``parameter space" of coupling constants. A 't Hooft anomaly in the space of couplings, following \cite{Seiberg:2018ntt,Cordova:2019jnf,Cordova:2019uob}, can be described as a failure of gauge-invariance for the partition function under a continuous  transformation of a scalar parameter. In particular,  if the partition function is multiplied by a background counterterm with a quantized coefficient, the phenomenon cannot be trivialized by making the coefficient parameter-dependent in a smooth way and thus it is interpreted as a novel example of 't Hooft anomaly. An important application of these ideas is in the context Yang--Mills theory with a $\theta$-angle \cite{Gaiotto:2017yup}. More generally, we can adopt such periodic scalar backgrounds to label families of quantum field theories.

In our analysis, we will focus on some simple examples of continuous $2$-group global symmetries that appear in the study of Goldstone models in three dimensions for a spontaneously broken $U(1)\zero_A$ symmetry. To make contact with the previous discussion, we also introduce a topological coupling to a background field $\theta$ that is circle-valued. Heuristically, we can view $\theta$ as a background field for a ``$(-1)$-form symmetry" $\bZ$ implemented by $2\pi$ shifts.\footnote{In this formulation, one can define $\theta$ by introducing a set of local coordinate patches such that the transition functions jumps by $2\pi\bZ$. The field strength $d\theta$ is however single valued and we can thus make sense of it within the background effective action. The proper mathematical formalism to discuss effective actions involving periodic scalar backgrounds is that of differential cohomology. We refer the interested reader to \cite{Cordova:2019jnf} for a rigorous account of this subject. } In all these examples, the change of the partition function $Z[A\one, \theta]$ under a background gauge transformation for $A\one$ can be associated with the following anomalous Ward identity
\be
d* j^{(1)}_A= \frac{k}{4\pi^2} d\theta\wedge F^{(2)}_A\ec \label{eq:0noncons}
\ee
with $k\in \bZ$ and $j_A\one$ denoting the $U(1)\zero_A$ current operator. Proceeding as in four dimensions, we can consider gauging the background field $\theta$, that is we replace $\theta\to \phi$ and integrate over it. In this context we can think of $\phi$ as an axion. As before, a consequence of this gauging procedure is that the right hand side of \eqref{eq:0noncons} should no longer be intepreted as a 't Hooft anomaly. The non-invariance under $U(1)\zero_A$ background gauge transformations can be addressed by noticing that, when $\phi$ is dynamical, there is a new conserved $2$-form current
\be
J\two_B = \frac{1}{2\pi}*d\phi\ed
\ee
together with its associated $2$-form background field $B\two$. The symmetry $U(1)_B\one$ can be coupled to the original theory as follows
\begin{align}
    \tilde Z[A^{(1)},B^{(2)}]\equiv\int {\cal D}\phi \ Z[A^{(1)},\phi] \ e^{\frac{i}{2\pi}\int B^{(2)}\wedge d\phi}\ed
\end{align}
Therefore, gauge invariance can be restored by imposing the following background gauge transformations
\begin{align}
A^{(1)} \to A\one + d\lambda_A^{(0)}\ec\quad B^{(2)}\to B\two + d\Lambda_B^{(1)} +  \frac{k}{2\pi} \lambda^{(0)}_A F_A^{(2)}\ed
\end{align}
signaling a continuous $2$-group global symmetry structure of the form
\be
U(1)^{(0)}_A\times_{\kappa=k} U(1)^{(1)}_B\ed
\label{eq:cc2gr}
\ee
As we will discuss in section \ref{sec:examples}, the procedure outlined here can be generalized in various directions to include a richer set of couplings and of global symmetry groups. We emphasize that similar manipulations are also available when the three-dimensional theory exhibit discrete global symmetry groups that can be studied by introducing discrete background gauge fields. 
A surprising outcome of our analysis is that gauging a compact background $\theta$ can also give rise to further subtle generalizations of the notion of global symmetry that go beyond $2$-group global symmetries. In section \ref{sec:GM}, we describe such phenomenon in a three-dimensional model with a $U(1)\zero_A$ Goldstone boson coupled topologically to a $U(1)\zero_c$ gauge field $c\one$. After gauging the compact background $\theta$ and replacing it by a dynamical $\phi$ one finds a non-conservation equation for the $U(1)\zero_A$ current that reads as follows
\be\label{ABJish}
d*j\one_A = \frac{k}{4\pi^2}d\phi\wedge f\two_c\ed
\ee
There is a fundamental difference between the characteristic $2$-group Ward identity \eqref{eq:0noncons} and \eqref{ABJish}. In the first case, when $F\two_A=0$, one has $d*j\one_A=0$ implying that the symmetry is conserved in absence of background fields at separated points. The right hand side of \eqref{ABJish} is instead \emph{operator-valued} and leads to a violation of the $U(1)\zero_A$ symmetry.\footnote{This can be thought of as an analogue of the ABJ anomaly in four dimensions.} In fact, for $k>1$, the global symmetry associated to $j_A^{(1)}$ is reduced to $\bZ_k^{(0)}$ and we will describe an example where this symmetry participates in a 2-group in section \ref{GM}. 

Quite surprisingly the value $k=1$ does not completely invalidate the $U(1)\zero_A$ symmetry. Following  \cite{Choi:2022jqy,Cordova:2022ieu}, we will show that in this case the three-dimensional Goldstone-Maxwell model admits an infinite set of topological operators $\cD_{\alpha}$ labelled by a rational number $\alpha \in \mathbb{Q}$ that can be obtained by dressing the $U(1)\zero_A$ symmetry defect. The topological operators $\cD_\alpha$ do not obey group law multiplication, in particular it is not possible to find an inverse operator $\cD^{-1}_\alpha$ such that $\cD_\alpha\times \cD^{-1}_\alpha = 1$. These are commonly referred to as \emph{non-invertible} global symmetries and have recently attracted a lot of attention in the literature \cite{Frohlich:2006ch,Bhardwaj:2017xup,Chang:2018iay,Thorngren:2019iar, Gaiotto:2020iye,Komargodski:2020mxz,Nguyen:2021yld,Thorngren:2021yso,Koide:2021zxj,Choi:2021kmx,Kaidi:2021xfk, Burbano:2021loy,Choi:2022zal,Roumpedakis:2022aik,Hayashi:2022fkw,Arias-Tamargo:2022nlf,Bhardwaj:2022yxj,Kaidi:2022uux,Antinucci:2022eat,Bashmakov:2022jtl}. Furthermore, an analogous infinite set of symmetry defects can be defined for any integer value of $k$, though through a slightly more involved mechanism which we will not study in detail in this paper.  We believe that the three-dimensional Goldstone--Maxwell model presented here can be useful to highlight various properties of non-invertible symmetries and their interplay with $2$-group global symmetries. 

The rest of the paper is organized as follows. In section \ref{sec:examples} we present a class of examples featuring $2$-group global symmetries in order to illustrate the general gauging procedure outlined above. All these examples involve background compact scalar fields whose gauging plays an essential role in the construction. In section \ref{sec:non-inv} we describe how the Goldstone--Maxwell model admits an infinite set of non-invertible global symmetries. In section \ref{sec:holo} we comment on holographic realizations of continuous $2$-group and non-invertible global symmetries in three dimensions by studying some simple bottom-up models of dynamical gauge fields in AdS$_4$. Finally, appendix \ref{apx} is devoted to a class of more exotic examples of $2$-groups.

\section{Continuous Higher Group Global Symmetries in Three Dimensions}\label{sec:examples}

\subsection{Goldstones Model}\label{1G}

A simple theory exhibiting most of the features of interest in this work is that of a single Goldstone boson  $\chi$ for a $U(1)^{(0)}_A$ global symmetry in three dimensions whose non-linear gauge transformation is:
\be
\chi \to \chi + \lambda^{(0)}_A\ed
\label{eq:Goldgauge}
\ee
In presence of a non-trivial background $A\one$ for $U(1)_A^{(0)}$, we can supplement the effective action by a term coupling $\chi$ to $F\two_A = {dA\one}$ as follows,
\be
S[A\one,\chi]=-\half\int (d\chi-A\one)\wedge* (d\chi-A\one) - \frac{i\theta}{4\pi^2}\int d\chi \wedge F\two_A\ed
\label{eq:GoldA}
\ee
Note that the $\theta$-angle has $2\pi$ periodicity following from the condition
\be
\int_{\Sigma_1}{d\chi\over2\pi}\in \bZ\ec\quad
\int_{\Sigma_2}{F\two_A\over2\pi} \in \bZ\ed
\ee
for any closed $1$-cycle $\Sigma_1$ and $2$-cycle $\Sigma_2$ in spacetime. The theory also admits a $1$-form global symmetry $U(1)\one_{\tilde B}$ whose conserved current is
\be
J\two_{\tilde B} = {1\over 2\pi}*d\chi\ed
\ee
The symmetry $U(1)\one_{\tilde B}$ is never spontaneously broken in three spacetime dimensions due to a generalized version of the classic Coleman--Mermin--Wagner theorem \cite{GKSW,Hofman:2018lfz,Lake:2018dqm}. We can couple it minimally to \eqref{eq:GoldA} as follows:
\be
\begin{split}
S[A\one,\tilde{B}\two, \chi]&=-\half\int (d\chi-A\one)\wedge* (d\chi-A\one) - \frac{i\theta}{4\pi^2}\int d\chi \wedge F\two_A\\ &+{i\over 2\pi}\int \tilde{B}\two \wedge d\chi\ed
\label{GoldAtB}
\end{split}
\ee
The above action is not invariant under the standard background gauge transformations $A\one \to A\one + d\lambda\zero_A$ and $\tilde{B}\two \to \tilde{B}\two + d\Lambda\one_{\tilde{B}}$. Thus, $U(1)_A\zero$ and $U(1)\one_{\tilde B}$ have a mixed 't Hooft anomaly whose inflow action is given by
\be \label{anomalyGoldstone}
S_{4d}[A\one,\tilde{B}\two] = -{i\over 2\pi}\int_{\cY_4}A\one\wedge d\tilde{B}\two\ec
\ee
where $\cY_4$ is a four-dimensional manifold whose boundary is the physical spacetime.

An important point, which will appear repeatedly below, is that $\theta$ can be promoted to a spacetime dependent background field. In particular, it can be made to depend on a single 
coordinate $x$. This introduces a winding around the circle as $x$ varies from $-\infty$ to $+\infty$. We will refer to such object as a ``background gauge field" for a $(-1)$-form symmetry. Note that the exterior derivative $d\theta$ is single valued and that is why we can extend the theory of ordinary anomalies to include such configurations as discussed in the introduction.

With this in mind, under a $U(1)_A^{(0)}$ gauge transformation \eqref{eq:Goldgauge} (ignoring the  't Hooft anomaly \eqref{anomalyGoldstone}), the action changes as 
\be\label{AISC1}
S[A\one +d\lambda\zero_A,\tilde{B}\two, \theta,\chi]-S[A\one,\tilde{B}\two, \theta, \chi]= \frac{i}{4\pi^2}\int \lambda\zero_A d\theta\wedge F\two_A\ec
\ee
hence leading to the anomalous conservation equation \eqref{eq:0noncons} (with $k=1$). 

At this point we can render $\theta$ dynamical by promoting $\theta\to \phi$, with $\phi$ a periodic scalar field with periodicity $2\pi$. As explained in the introduction, this gives rise to two emergent global symmetries. As before, there is a $U(1)\zero_{\tilde A}$ ordinary shift symmetry for $\phi$ and a $1$-form global symmetry $U(1)_B^{(1)}$ whose conserved current is given by:
\be\label{u1winding}
J\two_B = {1\over 2\pi}*d\phi\ed
\ee
The minimal coupling to backgrounds fields for all the global symmetries is given by:
\begin{align}
S[A\one,\tilde{A}\one, B\two,\tilde{B}\two,\chi,&\phi]=-\frac12\int (d\chi - A\one)\wedge* (d\chi- A\one) 
-\frac12\int  (d\phi - \tilde{A}\one)\wedge* (d\phi- \tilde{A}\one)\nonumber\\ &- \frac{i}{4\pi^2}\int \phi \,d\chi \wedge F_A\two +\frac{i}{2\pi}\int B\two\wedge d\phi +{i\over 2\pi}\int \tilde{B}\two \wedge d\chi\label{eq:2Gold}\ed
\end{align}
Even though $U(1)\zero_A$ participates in the anomaly \eqref{AISC1}, gauging of $\theta$ does not imply any inconsistency. Indeed, one can restore invariance under  $U(1)_A^{(0)}$ background gauge transformations \eqref{eq:Goldgauge}, by introducing the following compensating transformation rules
\be\label{2grouptrans}
B\two\to B\two + d\Lambda\one_B -\frac{1}{2\pi}\lambda\zero_A F\two_A\ec \quad  \tilde{B}\two\to \tilde{B}\two + d\Lambda\one_{\tilde{B}} + \frac{1}{2\pi}\lambda\zero_{\tilde{A}}F\two_A\ed
\ee
This implies that $U(1)_A\zero$, $U(1)_{\tilde A}\zero$, $U(1)_B\one$, $U(1)_{\tilde B}\one$ participate in a non-trivial 2-group global symmetry
\be
\left(U(1)^{(0)}_A\times U(1)^{(0)}_{\tilde{A}}\right)\times_{\kappa = 1} \left(U(1)^{(1)}_B\times U(1)_{\tilde B}\one\right)\ed
\ee
The complete four-dimensional inflow action for \eqref{eq:2Gold} is now given by
\be\label{adjustedinflow}
S_{4d}[A\one,\tilde{A}\one,\tilde{B}\two,B\two] = {i\over 2\pi}\int_{\cY_4}-A\one\wedge d\tilde{B}\two - \tilde{A}\one\wedge dB\two + {1\over 2\pi}A\one \wedge \tilde{A}\one\wedge F\two_A\ec
\ee
where the first two terms account for the mixed 't Hooft anomalies between $U(1)_{A,\tilde A}\zero$ and $U(1)\one_{\tilde{B},B}$ while the last term is a choice of counterterm to ensure invariance of the action \eqref{eq:2Gold} under the $2$-group background gauge transformations \eqref{2grouptrans}. We stress that such counterterm adjustments are a common requirement in models that exhibit non-trivial $2$-group global symmetries \cite{CDI,Benini:2018reh}.  

At this stage, it is interesting to make contact with the work \cite{Brauner:2020rtz} which focuses on the physics of higher-group global symmetries for effective actions of Goldstone modes in general space-time dimensions.  

In that context, the theory \eqref{eq:GoldA} is modified introducing an additional $U(1)\zero_C$ global symmetry with background gauge field $C\one$ which is topologically coupled to the Goldstone field $\chi$. The anomalous variation (ignoring ordinary 't Hooft anomalies) in presence of a background field $\theta$ is thus modified: 
\be\label{goldstonewithC}
S[A\one +d\lambda\zero_A,C\one, \tilde{B}\two, \theta,\chi]-S[A\one,C\one,\tilde{B}\two, \theta,\chi]= \frac{i}{4\pi^2}\int \lambda\zero_A d\theta\wedge F\two_C\ec
\ee
where $F\two_C=dC^{(1)}$.

The gauging of $\theta$ gives rise to a topological coupling of the form
\be
S[A\one,\tilde{A}\one, C\one, B\two,\tilde{B}\two,\chi,\phi] \supset  \frac{i}{4\pi^2}\int C\one\wedge d\phi\wedge d\chi\ec
\ee
which one can interpret as a minimal coupling to a composite topological current (sometimes referred to as a ``Chern--Weil" current \cite{Heidenreich:2020pkc}) of the form
\be
j\one_{\textrm{CW}} = {1\over 4\pi^2}*(d\phi \wedge d\chi)\ed
\ee
It is then easy to establish that the model discussed in \cite{Brauner:2020rtz} has a $2$-group involving
\be
\left(U(1)^{(0)}_A\times U(1)^{(0)}_{\tilde{A}} \times U(1)\zero_C\right)\times_{\kappa = 1} \left(U(1)^{(1)}_B\times U(1)_{\tilde B}\one\right)\ed \label{goldstone2group}
\ee
with $2$-group background gauge transformations and four-dimensional inflow action obtained with the same approach used in this section. 

To conclude this section, we would like to explain how to modify the gauging procedure of $\theta$ to obtain higher integer values $\kappa>1$ of the $2$-group structure constant. As remarked in the introduction, we view $\theta$ as a compact background field implementing a $\bZ$ global symmetry. We can always choose to gauge a subgroup of such symmetry and identify $\theta\sim \theta + 2\pi k$ with $k$ an arbitrary integer number, such choice amounts to gauging a $k$-fold cover of the symmetry $\bZ$. To implement such gauging procedure we substitute $\theta\to k \phi$, where $\phi$ is the dynamical axion introduced above. This allows to obtain a $2$-group global symmetry \eqref{goldstone2group} with structure constant $\kappa = k>1$.

\subsection{Goldstone--Maxwell Model} \label{sec:GM}

Consider the following theory consisting of a dynamical $U(1)$ gauge field denoted by $c\one$ and a Goldstone boson $\chi$, transforming non-linearly under a $U(1)\zero_A$ global symmetry, coupled with a $\theta$-term as follows:
\be\label{GM}
S[c\one, \chi]= -\frac{1}{2g^2}\int f\two_c\wedge* f\two_c - \frac12\int  d\chi\wedge* d\chi + \frac{i \theta}{4\pi^2}\int d\chi \wedge f\two_c \ec
\ee
with flux quantization chosen as in the previous section. Note that this is actually a gauging of the model discussed around \eqref{goldstonewithC}, by promoting $C^{(1)} \to c\one$. A similar model appears in the spontaneously broken phase of a $U(1)\zero \times_{\kappa} U(1)\one$ in four-dimensions \cite{CDI}.

Let us begin our analysis by listing here all the global symmetries of the Goldstone--Maxwell (GM) model. In three spacetime dimensions the Maxwell kinetic term gives rise to a magnetic $0$-form symmetry $U(1)\zero_{\text m}$ acting on 't Hooft monopole operators and
an electric $1$-form symmetry $U(1)_{\text e}\one$ acting on Abelian Wilson lines. In addition, the theory exhibits a $0$-form shift symmetry $U(1)_{A}\zero$ and winding $1$-form symmetry $U(1)\one_{\tilde B}$ due to the Goldstone field dynamics. The complete set of currents for these symmetries is given by
\be\label{currents}
j\one_A = -i d\chi\ec\qquad
J\two_{\text e}=-\frac{i}{g^2}f\two_c\ec\qquad
j_{\text m}\one = {1\over 2\pi}* f\two_c\ec \qquad  J\two_{\tilde B} = {1\over 2\pi}*d\chi  \ed
\ee
Both $j\one_A$ and $J\two_{\textrm{e}}$ could in principle be modified by an improvement term due to the $\theta$-term in \eqref{GM} which we prefer not to include in our presentation of the currents (though these improvements are important to find the relevant set of charges \cite{Witten:1979ey}). 

Following the discussion in section \ref{1G}, we can again promote $\theta$ to a background field in the effective action and gauge it in order to establish that the GM model exhibits a $2$-group structure. An important subtlety is that, in this example, one can retain a non-trivial 2-group structure only after gauging a subgroup of the $\bZ$-shift symmetry by identifying $\theta\sim \theta + 2\pi k$ with $k\in \bZ$. 

That is, if we replace $\theta$ by a compact dynamical field $\phi$ with $2\pi$ periodicity as in section \ref{1G} we must write the resulting gauged theory as follows, 
\be
S[c\one, \chi, \phi]=-\frac{1}{2g^2}\int f\two_c\wedge* f\two_c -\frac12\int  d\chi\wedge* d\chi- \frac12\int  d\phi\wedge* d\phi + \frac{i k}{4\pi^2}\int \phi d\chi \wedge f\two_c\ed 
\label{eq:2axionMaxwell}
\ee
Note that, gauging $\theta$ as above induces the following non-conservation equations derived from the expressions \eqref{currents} and the equations of motion of \eqref{eq:2axionMaxwell}
\be
d*j\one_A = -{k\over 4\pi^2} d\phi \wedge f\two_c\ec\qquad d*J_{\textrm{e}}\two =  -{k\over 4\pi^2} d\phi \wedge d\chi\ed \label{ABJnoncons}
\ee
The above conservation equations imply, using arguments similar to \cite{Brennan:2020ehu}, that the corresponding topological charge operators are gauge invariant only provided that both $U(1)\zero_A$ and $U(1)_{\textrm{e}}\one$ are explicitly broken to finite subgroups $\bZ\zero_{k,A}$ and $\bZ\one_{k,\textrm{e}}$.

It is natural to consider what happens if we instead decide to gauge the full shift symmetry implemented by $\theta$ by identifying $\theta \sim \theta + 2\pi$. In this case gauging would automatically violate both $U(1)\zero_A$ and $U(1)_{\textrm{e}}\one$ symmetries. Nevertheless, it is still possible to define an interesting set of topological operators that do not obey a group multiplication law but rather a \emph{non-invertible} fusion algebra. We will further expand on this point in section \ref{sec:non-inv}.

We are finally ready to describe the full list of global symmetries for the three-dimensional Goldstone--Maxwell model \eqref{eq:2axionMaxwell} which includes   
\be
\mathbb{Z}_{k,A}^{(0)}\times \mathbb{Z}_{k,\tilde{A}}^{(0)}\times U(1)^{(0)}_{\textrm{m}} \times  \mathbb{Z}_{k,\textrm{e}}^{(1)}\times U(1)^{(1)}_{\tilde B}\times U(1)^{(1)}_{B}\ec
\label{eq:2axionG}
\ee
where the second and the last factor are emergent global symmetries induced by gauging $\theta$ as we described in the previous example, and furthermore $\mathbb{Z}_{k,\tilde{A}}^{(0)}$ results from the breaking of a $U(1)^{(0)}_{\tilde{A}}$ by the same mechanism described after \eqref{ABJnoncons}, but applied to $j_{\tilde A}\one=-id\phi$.

The Goldstone--Maxwell model has a $2$-group symmetry structure which we can obtain by coupling the theory to a general configuration of background fields associated to the whole set of global symmetries \eqref{eq:2axionG}.\footnote{A $\Z^{(p)}_k$ background field is defined by a $(p+1)$-cocycle $\omega^{(p+1)}\in H^{p+1}(\cM_3,\Z_k)$. Throughout this work we represent these objects in terms of flat $U(1)$ $(p+1)$-form gauge fields $A^{(p+1)}$ whose gauge equivalence classes $[A^{(p+1)}]$ satisfy $[A^{(p+1)}] = \frac{2\pi}{k}\omega^{(p+1)}$.} Related higher group global symmetries are also present in four-dimensional axion models \cite{Brennan:2020ehu,Hidaka:2020iaz,Hidaka:2021mml}.

Following \cite{Brennan:2020ehu}, we can extend the topological coupling of Goldstones  \eqref{eq:2axionMaxwell} to a four dimensional bulk manifold $\cY_4$, such that $\partial \cY_4=\cM_3$ with $\cM_3$ the physical three-dimensional spacetime manifold
\be
\frac{i k}{4\pi^2}\int_{\cM_3}\phi d\chi\wedge f\two_c= \frac{i k}{4\pi^2}\int_{\cY_4}d\phi \wedge d\chi\wedge f\two_c
\ee  
The physical theory on $\cM_3$ cannot depend on the choice of bulk extension, that is if we choose two different extension manifolds $\cY_4$ and $\cY_4'$ we must require that 
\be
\frac{i k}{4\pi^2}\int_{\cY_4\cup_{\cM_3} \bar \cY'_4}d\phi \wedge d\chi\wedge f\two_c= 2\pi i \mathbb{Z}
\label{eq:cons}
\ee 
where $\cup_{\cM_3}$ denotes gluing along the boundary $\cM_3$ and $\bar{\cY}'_4$ is the orientation reversal of $\cY'_4$. The above consistency condition is clearly satisfied iff $k\in \mathbb{Z}$. The action \eqref{eq:2axionMaxwell} extended to $\cY_4$ and coupled to the relevant background fields includes terms like
\be
\begin{split}
   S[A\one, \tilde{A}\one, A_{\textrm{m}}\one, B\two_{\textrm{e}},B\two,\tilde{B}\two,c\one, \chi, \phi]&\supset
\frac{i k}{4\pi^2}\int_{\cY_4}  (d\phi-\tilde A\one) \wedge (d\chi- A\one) \wedge (f_c\two-B\two_{\textrm{e}}) \label{eq:2axionBack}\\
&+ \frac{i}{2\pi}\int_{\cY_4} H^{(3)}_{B} \wedge d\phi +  H^{(3)}_{\tilde{B}} \wedge d\chi + F\two_{\textrm{m}} \wedge f\two_c \ec
\end{split}
\ee
where we have written the field strengths for the background fields of the magnetic symmetries.

To ensure that the effective action, as a function of all the background fields, does not depend on the choice of bulk extension $\cY_4$ we must enforce the condition \eqref{eq:cons} which requires
\be\label{bulkextension}
\begin{split}
\frac{i}{2\pi}\int_{\cY_4\cup_{\cM_3} \bar \cY'_4}&\left[ \left(H^{(3)}_{B} -\frac{k}{2\pi}A\one \wedge B\two_{\textrm{e}}\right)\wedge d\phi + \left( H^{(3)}_{\tilde{B}} +\frac{k}{2\pi}\tilde{A}\one \wedge B\two_{\textrm{e}}\right) \wedge d\chi\right.\\  &\left.+\left(F\two_{\textrm{m}}+\frac{k}{2\pi}\tilde{A}\one \wedge A\one\right) \wedge f\two_c \right] = 0\mod  2\pi i \mathbb{Z}\ed
 \end{split}
\ee
In order for \eqref{bulkextension} to be satisfied, 
we see that the background field strengths must be modified, so that the combinations appearing above become exact and hence integrally quantized
\be
H^{(3)}_{B}= dB\two +\frac{k}{2\pi}A\one \wedge B\two_{\textrm{e}}\ec \quad H^{(3)}_{\tilde{B}}=d\tilde{B}\two-\frac{k}{2\pi}\tilde{A}\one \wedge B\two_{\textrm{e}}\ec\quad \label{eq:2grFS}
F\two_{\textrm{m}}=dA\one_{\textrm{m}} -\frac{k}{2\pi}\tilde{A}\one \wedge A\one\ed
\ee
Therefore, we conclude that the global symmetries of \eqref{eq:2axionMaxwell} participate in a $2$-group global symmetry which we denote by  
\be\label{2groupGMfinal}
\left(\mathbb{Z}_{k,A}^{(0)}\times \mathbb{Z}_{k,\tilde{A}}^{(0)}\times \mathbb{Z}_{k,\textrm{e}}^{(1)}\right) \times_{\kappa = k} \left(U(1)^{(0)}_{\textrm{m}}\times U(1)^{(1)}_{\tilde B}\times U(1)^{(1)}_{B}\right)\ec
\ee
with $2$-group background fields gauge transformations given by 
\be\label{2groupGMtransf}
\begin{split}
&A\one_{\textrm{m}}\to A\one_{\textrm{m}} + d\lambda\zero_{\textrm{m}} + \frac{k}{2\pi}\lambda\zero_{\tilde{A}} \wedge A\one-\frac{k}{2\pi}  \lambda\zero_{A}\wedge \tilde{A}\one+\frac{k}{2\pi} \lambda\zero_{\tilde{A}} \wedge d\lambda\zero_{A}\ec\\
&B\two\to B\two + d\Lambda\one_B - \frac{k}{2\pi} \lambda\zero_{A}\wedge  B\two_{\textrm{e}}-\frac{k}{2\pi} \Lambda\one_{\textrm{e}}\wedge A\one-\frac{k}{2\pi} \lambda\zero_{A}\wedge  d\Lambda\one_{\textrm{e}}\ec \\
&\tilde{B}\two\to \tilde{B}\two + d \Lambda\one_{\tilde{B}} + \frac{k}{2\pi} \lambda\zero_{\tilde{A}}\wedge  B\two_{\textrm{e}}+\frac{k}{2\pi} \Lambda\one_{\textrm{e}}\wedge \tilde{A}\one+\frac{k}{2\pi} \lambda\zero_{\tilde{A}}\wedge  d\Lambda\one_{\textrm{e}}\ed
\end{split}  
\ee
Note that in \eqref{2groupGMfinal} we have a slightly unconventional occurrence of a $0$-form symmetry on the right of the $2$-group product, and similarly a discrete $1$-form symmetry on its left. Our convention is that we write on the left all the symmetries whose background fields modify the background transformation laws of the symmetries that appear on the right.\footnote{A further remark is that the first line of \eqref{2groupGMtransf} has a $2$-group-like structure, but involves only $0$-form symmetries. However, since the symmetries modifying the transformation law are discrete, this is a genuinely non-trivial structure that cannot be trivialized by a field redefinition.}

Finally, we can write down the complete four-dimensional inflow action for \eqref{eq:2axionMaxwell} which is given by 
\be\label{GM2grouptransf}
\begin{split}
S_{\textrm{4d}}[A\one, \tilde{A}\one, A_{\textrm{m}}\one, B\two_{\textrm{e}},B\two,\tilde{B}\two] = -\frac{i}{2\pi}\int_{\cY_4}A\one\wedge  \tilde H^{(3)} &+\tilde A\one\wedge  H^{(3)} -B\two_{\rm e}\wedge F\two_m \\ &+\frac{k}{2\pi}A\one\wedge \tilde A\one\wedge  B\two_{\rm e}
\end{split}
\ee
In the above expression, the first two terms are mixed anomalies, induced by the Goldstones, between respectively $\bZ\zero_{k,A}$ and $U(1)\one_{\tilde{B}}$, and $\bZ\zero_{k,\tilde{A}}$ and $U(1)\one_{B}$.  The third term signals a mixed 't Hooft anomaly between $\bZ_{k,\textrm{e}}\one$ and $U(1)_{\textrm{m}}\zero$, from the Maxwell term. Finally, the last term is a counterterm adjustment which is needed to ensure gauge invariance under the background gauge transformations introduced in \eqref{eq:2grFS}.

\subsection{Goldstone--Yang--Mills Model}

The previous model allows for a natural generalization with non-Abelian gauge fields. For simplicity, we will only consider a $SU(N)$ gauge group even though our analysis can be repeated for any simply connected group. The resulting action is
\be\label{GYM}
S[a\one, \chi]=-\frac{1}{2g^2}\int {\rm Tr}(f_a\two\wedge* f_a\two) -\frac12\int {\rm Tr}(D\chi\wedge* D\chi) + \frac{i \theta}{4\pi^2}\int {\rm Tr}(D\chi \wedge f\two_a)\ec
\ee
where $a\one$ is a $SU(N)$ gauge field whose field strength is $f\two_a = da\one + a\one \wedge a\one$ and $\chi$ is a compact scalar also transforming in the adjoint representation of $SU(N)$ whose covariant derivative is $D\chi = d\chi + [a\one, \chi]$. Note that it is necessary for $\chi$ to transform in the adjoint representation for the above $\theta$-term to make sense.\footnote{A similar term was also discussed in \cite{BenettiGenolini:2020doj}.}

The global symmetries of \eqref{GYM} include an ordinary $0$-form shift symmetry $\mathbb{Z}_{N,A}^{(0)}$ acting on the operator $U = e^{i \chi(x)}$ as $U \to e^{2\pi i \over N}U$. Since all the fields transform in the adjoint representation, the system also admits a $1$-form global symmetry $\mathbb{Z}_{N}^{(1)}$ acting on fundamental Wilson lines. Whenever both background gauge fields $A\one$ and $B_{\textrm{e}}\two$ are activated it is possible to show that \eqref{GYM} exhibits an anomaly in the space of couplings of the form\footnote{This result follows from the modified quantization condition of the topological coupling in \eqref{GYM} whenever both background gauge fields $A\one$ and $B_{\textrm{e}}\two$ are activated:
\be
\frac{1}{4\pi^2}\oint{\rm Tr}\left(D\chi \wedge f\two_a\right)= \frac{N(N-1)}{4\pi^2}\oint A\one\wedge B_{\textrm{e}}\two \mod\mathbb{Z}\ed
\label{eq:fracChern}
\ee
}
\be
S[A\one + d\lambda\zero_A, B_{\textrm{e}}\two, \theta, a\one, \chi]-S[A\one, B_{\textrm{e}}\two, \theta, a\one, \chi]= - i \frac{N(N-1)}{4\pi^2}\int \lambda\zero_A d\theta\wedge B_{\textrm{e}}\two\ed
\label{eq:mixNonAbelian3d}
\ee  
Therefore, proceeding as in the previous example, we can gauge  a subgroup of the shift symmetry implemented by $\theta$ and obtain
\be
S[a\one, \chi, \phi]= -\frac{1}{2}\int {\rm Tr}(f\two_a\wedge* f\two_a) - \frac12 \int{\rm Tr}(D\chi\wedge* D\chi)-\frac12\int d\phi\wedge* d\phi + \frac{i k}{4\pi^2}\int \phi{\rm Tr}(D\chi \wedge f\two_a)\ed
\label{eq:2axionYM}
\ee
We will now show that this theory exhibits a $2$-group global symmetry as a direct consequence of the anomaly \eqref{eq:mixNonAbelian3d}. The relevant global symmetries to discuss for this purpose are 
\be
\mathbb{Z}_{N,A}^{(0)}\times \mathbb{Z}^{(1)}_{N,B_{\textrm{e}}} \times U(1)^{(1)}_{B}\ec 
\label{eq:2axionYMG}
\ee
where the last factor is a $U(1)\one$ winding symmetry whose current $J\two_B$ was already introduced in \eqref{u1winding}.

The fully coupled theory only makes sense if we  enforce a quantization condition on its bulk extension \be
\frac{i}{2\pi}\int_{\cY_4\cup_{\cM_3} \bar \cY'_4} \left(H^{(3)}_{B} -\frac{kN(N-1)}{2\pi}A\one\wedge B\two_{\textrm{e}} \right)\wedge d\phi  = 0\mod  2\pi i \mathbb{Z}\ec
\ee
where $\cM_3$ is the physical spacetime and $\partial\cY_4 = \cM_3$. It is now clear that one needs to introduce a modified three-form field strength of the form
\be
H^{(3)}_B= dB\two + \frac{kN(N-1)}{2\pi}A\one\wedge B\two_{\textrm{e}}\ec
\ee
signaling the presence of a $2$-group global symmetry
\be
\left(\mathbb{Z}_{N,A}^{(0)}\times \mathbb{Z}_{N,B_{\textrm{e}}}^{(1)}\right) \times_{\kappa = k} U(1)^{(1)}_{B}\ed
\ee

The construction presented in this section can be easily generalized to include variants of the theory which (for example) involve reduced discrete symmetries or matter fields in various representations for which one can consider finite bundles for the faithful $0$-form global symmetry.

\subsection{Dimensional Reduction from Four Dimensions}

Theories with periodic scalars such as the ones studied in previous examples arise naturally in the context of dimensional reduction. Intuitively, the compactness of scalar fields in $3d$ may be understood as a consequence of gauge invariance in the $4d$ theory. This connection can be made more explicit as follows. Consider a four-dimensional theory exhibiting a $U(1)\zero_A \times U(1)_C\zero$ global symmetry whose 't Hooft anomaly coefficients are such that $k_{C^3}=0$ and $k_{AC^2}=0$. A simple example of such theory has been already discussed in the introduction. The relevant five-dimensional anomaly inflow action is given by
\be\label{i6mixed}
\cI_{\textrm{mixed}}^{(5)}= -\frac12{k_{A^2C}\over (2\pi)^2}\int A\one\wedge F\two_A\wedge F\two_C\ed
\ee
As it is well known, gauging the symmetry $U(1)_C\zero \to U(1)\zero_c$ leads to a 2-group global symmetry 
\be\label{4dtwogroup}
U(1)\zero_A \times_{\kappa = -{1\over 2}k_{A^2C}} U(1)\one_B\ed\ee 
It has also been discussed in \cite{CDI} that upon dimensional reduction, the $4d$ theory with symmetry \eqref{4dtwogroup} reduces to a three-dimensional theory with the same $2$-group global symmetry plus a new $0$-form symmetry $U(1)\zero_{\tilde B}$
\be\label{3dtwogrp}
\left(U(1)\zero_A \times_{\kappa = -{1\over 2}k_{A^2C}} U(1)\one_B\right) \times U(1)\zero_{\tilde B}\ed
\ee
The origin of $U(1)\zero_{\tilde B}$ can be traced by analyzing the dimensional reduction for the background gauge field transformations of \eqref{4dtwogroup}. 

We would now like to demonstrate that the symmetry structure \eqref{3dtwogrp} can also be understood in terms of the results presented in this work. The new point of view, following \cite{Ohmori:2021sqg}, is that we first perform a dimensional reduction of the action \eqref{i6mixed}. We integrate on a circle with local coordinate $\tau$, and further restrict $C\one$ to be along that circle, to obtain
\be\label{4dspt}
\cI_{\textrm{mixed}}^{(4)}= -\frac12{k_{A^2C}\over (2\pi)^2}\int A\one\wedge F\two_A\wedge d\theta\ec
\ee
where $A\one$ is now a three-dimensional background gauge field and $\theta \equiv C_\tau\one$ is a compact background scalar. 
The $4d$ invertible field theory \eqref{4dspt} can be thought of as an anomaly in the space of couplings between the periodicity of $\theta$ and the $U(1)_A\zero$ symmetry. It is not immediately obvious how such terms can arise purely in a three-dimensional effective theory. To see that, one should compute the contribution to the background effective action upon integrating over the whole tower of KK modes generated by the $4d$ massless Weyl fermions \cite{DiPietro:2014bca}.

We can now gauge $\theta$ replacing it by a dynamical compact scalar $\phi$, this introduces two new symmetries: a $0$-form shift symmetry $U(1)\zero_{\tilde A}$ and a $1$-form winding symmetry $U(1)_{B}\one$. We can furthermore modify  the background gauge transformation for $U(1)\one_B$ in the usual way to obtain a $2$-group global symmetry of the form
\be
\left(U(1)\zero_A \times_{\kappa = -{1\over 2}k_{A^2C}} U(1)\one_B\right) \times U(1)\zero_{\tilde A}\ec
\ee
which is exactly the same as \eqref{3dtwogrp} upon the identification of background fields $\tilde{B}\one = \tilde{A}\one$.

\section{Non-Invertible Symmetries in the Goldstone--Maxwell Model}\label{sec:non-inv}
In this section we discuss a further generalization of the notion of global symmetry emerging from the analysis of the three-dimensional Goldstone--Maxwell model. 

As we shown in section \ref{sec:GM}, the action \eqref{eq:2axionMaxwell} leads to the following non-conservation equations\footnote{The theory \eqref{eq:2axionMaxwell} also has a non-conserved $U(1)\zero_{\tilde{A}}$ symmetry whose current $j\one_{\tilde A}$ obeys a similar equation as $j\one_A$ in \eqref{nonconv} where one replaces $\phi$ with $\chi$. Note that, all the results regarding $U(1)\zero_A$ contained in this section apply verbatim also to $U(1)\zero_{\tilde A}$.}
\be\label{nonconv}
d*j\one_A = {-}{k\over 4\pi^2} d\phi \wedge f\two_c\ec\qquad d*J_{\textrm{e}}\two =  {-}{k\over 4\pi^2} d\phi \wedge d\chi\ed
\ee
Moreover, for any integer $k>1$, the three-dimensional GM model exhibits a non-trivial $2$-group global symmetry structure. For $k=1$, there is no non-trivial discrete subgroup of $U(1)\zero_A$ and $U(1)\one_{{\textrm{e}}}$ which is left behind. Gauging the full shift symmetry $\theta \sim \theta +2 \pi$  of the background $\theta$ parameter in the parent theory \eqref{GM} (and not just a subgroup of it, $\theta \sim \theta + 2\pi k$) does not lead to a $2$-group global symmetry. What happens for $k=1$ is more subtle and has been recently discussed in four dimensions \cite{Choi:2022jqy,Cordova:2022ieu}. As we will soon see, in this case one has \emph{non-invertible} symmetry defects. 

To see how these objects come about in the three-dimensional GM model we will insist on defining the extended symmetry operators which would implement the symmetries $U(1)\zero_A$ and $U(1)\one_{\textrm{e}}$
\be
U_{\alpha}(\Sigma_2)=\exp\left(i\alpha\int_{\Sigma_2}*j\one_A\right)\ec \quad U_{\beta}(\Sigma_1)=\exp\left(i\beta\int_{\Sigma_1}*J\two_{\textrm{e}}\right)\ec
\ee
where $\Sigma_p$ are $p$-dimensional closed oriented submanifolds in spacetime and $\alpha,\beta \in [0, 2\pi)$. The problem with the extended operators $U_\alpha(\Sigma_2)$ and $U_\beta(\Sigma_1)$ is that they are clearly not topological in the sense of \cite{GKSW} since the corresponding currents are not conserved. We could try to remedy this problem by introducing a new set of extended operators as follows
\be
\hat{U}_{\alpha}(\Sigma_2)=\exp\left(i\alpha\int_{\Sigma_2}*j\one_A {+} {1\over 4\pi^2}\phi f\two_c\right)\ec \quad \hat{U}_{\beta}(\Sigma_1)=\exp\left(i\beta\int_{\Sigma_1}*J\two_{\textrm{e}} {+}{1\over 4\pi^2}\phi d\chi\right)\ec
\ee
which are obtained by modifying the currents $j\one_A\to\hat{j}\one_A$ and $J\two_{\textrm{e}}\to\hat{J}_{\textrm{e}}\two$ in such a way that formally $d*\hat{j}\one_A=0$ and $d*\hat{J}_{\textrm{e}}\two=0$. Even though $\hat{U}_{\alpha}(\Sigma_2)$ and $\hat{U}_{\beta}(\Sigma_1)$ are now topological, they are not gauge invariant operators. Indeed, both the $2d$ BF theory $\phi f\two_c$ and the $1d$ BF theory $\phi d\chi$ have an improperly quantized coupling. See for example the general discussion on $d$-dimensional BF theories in \cite{Kapustin:2014gua}.
The final step in this procedure is to consider whether there exist values of $\alpha$ or $\beta$ that allow to write a gauge invariant and topological extended operator. For example, when $\alpha = {2\pi \over N}$ and $N\in \bZ$, the operator $\hat{U}_{{2\pi\over N}}(\Sigma_2)$ can be replaced by the following extended operator
\be
\cD_{{1\over N}}(\Sigma_2)= \exp\left(i\int_{\Sigma_2} {2\pi\over N}*j\one_A + {N\over 2\pi}\xi dv\one + {1\over 2\pi}\xi f\two_c  {-} {1\over 2\pi}\phi dv\one \right)\ec
\ee
where $\xi$ is a compact scalar field and $v\one$ a 1-form gauge field, both defined uniquely on $\Sigma_2$, A path integral over both these fields is implicit in the expression above. The extended operator $\cD_{{1\over N}}(\Sigma_2)$ is defined by dressing the non gauge-invariant operator $U_{{2\pi\over N}}(\Sigma_2)$ by a $2d$ BF topological quantum field theory coupled to the $3d$ bulk fields $\phi$ and $f\two_c$. Similarly, we can repeat the entire operation with $\beta={2\pi\over M}$, $M\in \bZ$ and obtain
\be
\cD_{1\over M}(\Sigma_1)=\exp\left(i\int_{\Sigma_1} {2\pi\over M}*J\two_{\textrm{e}} + {M\over 2\pi}\xi d\tilde{\chi} + {1\over 2\pi}\xi d\chi {-}{1\over 2\pi}\phi d\tilde{\chi} \right)\ec
\ee
this time with $\xi$ and $\tilde{\chi}$ taken as compact scalars defined on $\Sigma_1$. Note that both the $2d$ and $1d$ BF theories can be understood as sequential reductions of the $3d$ theory for the Fractional Quantum Hall state.

The fusion algebra of the operators $\cD_{1\over N}$ and $\cD_{1\over M}$ interestingly does not follow the usual group-like relations. For example, consider the operator
\be
\cD^{\dagger}_{{1\over N}}(\Sigma_2)= \exp\left(-i\int_{\Sigma_2} {2\pi\over N}*j\one_A + {N\over 2\pi}\bar{\xi} d\bar{v}\one + {1\over 2\pi}\bar{\xi} f\two_c  {-}{1\over 2\pi}\phi d\bar{v}\one \right)\ec
\ee
with $\bar{\xi}$ and $\bar{v}\one$ defined only on $\Sigma_2$. The parallel fusion\footnote{A comprehensive study of the fusion algebra rules of $\cD_{1\over N}$ and $\cD_{1\over M}$ is left to future work. } of $\cD^{}_{{1\over N}}$ and $\cD^{\dagger}_{{1\over N}}$ leads to
\be
\begin{split}
\cD_{1\over N}(\Sigma_2) \times \cD^{\dagger}_{{1\over N}}(\Sigma_2) = \exp\left(i\int_{\Sigma_2}  {N\over 2\pi}\left(\xi dv\one -\bar{\xi} d\bar{v}\one\right) + {1\over 2\pi}\left(\xi-\bar{\xi}\right) f\two_c {-} {1\over 2\pi}\phi d\left(v\one-\bar{v}\one\right) \right)\ec
\end{split}
\ee
where the right hand side of this relation is called a condensation defect, which is not the identity. Indeed, it is the path integral over the direct product of two conjugate topological theories, which couple (anti)diagonally to the $3d$ fields as backgrounds. The above condensation defect can also be understood as a higher gauging of $\mathbb{Z}_N$ subgroups of the bulk magnetic and winding symmetries along $\Sigma_2$. We refer the interested reader to \cite{Roumpedakis:2022aik} where higher gaugings have been introduced. 

Of course, similar conclusions can be also reached for the operator $\cD_{1\over M}$. The topological operators $\cD_{1\over N}$ and $\cD_{1\over M}$ should be viewed as a further generalization of the notion of global symmetry with no reference to a symmetry group action. In the literature these are commonly referred to as non-invertible symmetry defects and have recently received a considerable amount of interest, see the introduction for a list of recent works on the subject. 

As emphasized in \cite{Choi:2022jqy, Cordova:2022ieu}, even though the set of non-invertible symmetries described here are discrete they are labeled by rational numbers which are dense in $\bR/2\pi\bZ$. A natural question is whether these generalized symmetries lead to novel physical implications for the Goldstone--Maxwell model. Finally, it is worth mentioning that these non-invertible defects also exists for $k>1$ where the GM model also exhibits a $2$-group global symmetry. It would be very interesting to understand what kind of generalized symmetry structure can be defined in this case.

\section{Comments on the Holographic Dictionary}\label{sec:holo}

In this section we describe how continuous $2$-group global symmetries can be understood from a holographic point of view. The role of higher-form symmetries in bottom-up models of holography has been recently addressed in several works (see for instance \cite{Hofman:2017vwr,Grozdanov:2017kyl,Iqbal:2020lrt,Iqbal:2021tui}).  Here we are only interested in characterizing the global symmetries featured in the putative dual field theory. For our purposes it is then enough to focus on a simple bottom-up setup describing gauge fields in AdS$_4$. A similar analysis has been recently presented in \cite{DeWolfe:2020uzb} for the case of AdS$_5$.

An important point to keep in mind when dealing with such holographic realizations is the following. A unitary conformal field theory in three-dimensions cannot host a continuous 1-form symmetry since the existence of a conserved current operator of dimension $\Delta_{J\two}=1$ is incompatible with the unitarity bounds on local primaries arising in the $3d$ conformal algebra \cite{Lee:2021obi}. A three-dimensional theory featuring a continuous $1$-form symmetry can still flow to an interacting CFT in the IR, provided that the symmetry decouples at sufficiently low energies. Accordingly, the effective theory hosting a non-trivially realized (continuous) 1-form symmetry is still a perfectly good description of the dynamics occurring at intermediate energy scales.

In holography these facts become manifest in a fairly simple way. A continuous $1$-form symmetry corresponds to a bulk \emph{dynamical} $U(1)$  $2$-form gauge field $b^{(2)}$. In three dimensions, corresponding to a bulk AdS$_4$ geometry, the electromagnetic dual of $b^{(2)}$ is a compact massless scalar $\chi$. 

As we will explain below, the dynamical field $\chi$ stands as a natural bulk realization of the compact scalar backgrounds fields that we used throughout this work. Now, being massless, it certainly lies outside the Breitenlohner-Freedman window $-9/4<m^2<-5/4$ \cite{Breitenlohner:1982bm}. The standard quantization for $\chi$ in AdS$_4$ is thus the only consistent choice that is available. Indeed, imposing the alternative quantization for $\chi$ leads to a violation of the unitarity bound ({\it c.f.} \cite{Andrade:2011dg}). 

It is well known that the alternative quantization of $\chi$ corresponds to the standard quantization of $b\two$. As we will see in the example below, the equations of motion for a massless $U(1)$ $2$-form gauge field in AdS$_4$ lead to a linearly divergent leading mode. We conclude that the dual field theory is not conformal. This is precisely how one can interpret holographically the absence of continuous $1$-form symmetries in three-dimensional conformal field theories.

Let us consider a rigid bulk AdS$_4$ geometry with $R_{\textrm{AdS}}=1$ whose metric, in Poincar\'e coordinates, reads
\be
ds^2 = \frac{dr}{r^2} + r^2 dx_idx^i\ec
\ee 
with conformal boundary placed at $r\to\infty$. In all the computations below, we adopt a radial gauge for both $U(1)$ gauge fields $a\one$ and $b\two$ by setting $a_{r}=b_{ri}=0$.

We begin our discussion with a four-dimensional bulk action depending on the following dynamical fields
\be
S_{\textrm{bulk}}[a\one, \chi] = \int   - \frac{1}{2}d\chi \wedge* d\chi -\frac{1}{2} f\two_a\wedge* f\two_a +\frac{k}{4\pi^2} d\chi \wedge a\one \wedge f\two_a\ec
\label{eq:holoaction}
\ee
with $k\in \mathbb{Z}$. We have not included coupling constants in front of the kinetic terms because they will not play any relevant role here. \footnote{Throughout this section we work in Lorentzian signature.}

For $k=0$, the solutions to the free field equations of motion have the following asymptotic form
\begin{align}
&\chi =\alpha + \frac{\gamma}{r^2} + \frac{\beta}{r^3}+\ldots\ec~~\gamma=\frac12 \partial^2 \alpha\ec \label{eq:Cfree}\\
&a_i = \alpha_i + \frac{\beta_i}{r}+\frac{\gamma_i}{r^2} + \ldots\ec~\gamma_i=\frac12 \partial_j h_{ij} \ec\quad \partial_i\beta_i=0\ec \label{eq:Afree}
\end{align}
with $h_{ij}\equiv \partial_{i}\alpha_{j}-\partial_{j}\alpha_{i}$. As usual, within the standard quantization, the leading component $\alpha_i$ in \eqref{eq:Afree} acts as a source coupled to a $U(1)\zero$ global symmetry conserved current in the boundary theory. The latter is realized by the subleading coefficient $j^{(1)}\equiv \beta_i$, and its conservation is imposed by the radial constraint. On the other hand, the leading coefficient $\alpha$ in \eqref{eq:Cfree} can now be interpreted as a compact scalar background at the boundary. Note that there is no additional constraint imposed on the subleading coefficient $\beta$. This is consistent with the absence of a conserved current associated to these symmetries.

In presence of a Chern--Simons-like interaction, the asymptotic structure of the solutions remains unchanged, however some relations between the coefficients are modified, most notably we now have
 \be
\partial_i \beta_i = -\frac{k}{4\pi^2}\epsilon^{ijk}\partial_i \alpha h_{jk}\ec\label{eq:coeffrel}
\ee
where we have used the sign convention $\epsilon^{ijkr}=\epsilon^{ijk}$ for the Levi-Civita tensor densities.

The on-shell action is divergent and one needs to regularize it by adding suitable boundary terms at a given cut-off radial position. We will not go into the details of the computation but just limit ourselves to list the final result for the renormalized boundary variation
\be
\delta S_{\textrm{bulk}}[a\one, \chi]+\delta S_{\textrm{c.t.}}[a\one, \chi]\Big|_{\textrm{on-shell}}=\int_{\textrm{bdry}} 3\delta\alpha\beta +\delta\alpha_i \beta_i - \frac{ k}{8\pi^2}\epsilon^{ijk}\left(\delta\alpha \alpha_i h_{jk}+2\delta\alpha_i\partial_j\alpha \alpha_k\right)\ec
\ee
from which we can readily recognize the current coupled to the $1$-form background
\be
j_i = \beta_i-\frac{ k}{4\pi^2}\epsilon^{ijk}\partial_j \alpha \alpha_k\ed
\ee
Finally, using \eqref{eq:coeffrel} we find the corresponding anomalous Ward identity
\be
\partial_i j_i =-\frac{k}{8\pi^2}\epsilon^{ijk}\partial_i \alpha h_{jk}\ec
\label{eq:holoward2}
\ee
which reproduces exactly \eqref{eq:0noncons}.

It is well known that gauging a global symmetry is interpreted holographically as a mapping from Dirichlet to free (Neumann) boundary conditions for the associated bulk gauge field \cite{Witten:2003ya,Witten:1998wy,Marolf:2006nd} (see \cite{Argurio:2022vtz} for a recent application). Equivalently, one may perform a Hodge duality and keep imposing Dirichlet boundary conditions on the dual gauge field. We will follow the latter procedure below. 

To dualize $\chi$ we introduce a Lagrange multiplier $2$-form $b\two$ in order to impose the Bianchi identity
\be
S_{\textrm{bulk}}[b\two, a\one, \chi] = \int   - \frac{1}{2}d\chi \wedge* d\chi -\frac{1}{2} f\two_a\wedge* f\two_a +\frac{k}{4\pi^2} d\chi \wedge a\one \wedge f\two_a - \frac{1}{2\pi}db\two\wedge d\chi\ed
\label{eq:hodge}
\ee
As a result, the dual action is simply\footnote{Recall that the Hodge operator when acting on a $p$-form, satisfies $*^2=(-1)^{s+p(d-p)}$ where $d$ is the spacetime dimension and $s=0$ ($s=1$) for Euclidean (Lorentzian) signature.}
\be
S_{\textrm{dual}}[b\two, a\one]= \int  - \frac{1}{2}c^{(3)} \wedge* c^{(3)} -\frac{1}{2} f\two_a\wedge* f\two_a\ec \label{eq:holdualaction}
\ee
where
\be
*d\chi = \frac{1}{2\pi}db\two+\frac{k}{4\pi^2} a\one\wedge f\two_a\equiv  c^{(3)}\ed \label{eq:duality}
\ee
The modified field strength $c^{(3)}$ satisfies
\be
dc^{(3)} = \frac{k}{4\pi^2}f\two_a\wedge f\two_a\ec
\ee
hence, the boundary $2$-group global symmetry is already manifest in the bulk due to a Green--Schwarz mechanism. 

According to the discussion at the beginning of this section, a continuous 1-form symmetry is not consistent with conformal invariance in $d=3$. This pathology manifests itself in the dual picture through a linearly divergent piece in the solution for the $2$-form free equations of motion
\be
b_{ij}=\alpha_{ij}+r \beta_{ij} + \frac{\gamma_{ij}}{r} + \ldots\ec \quad \partial_j\beta_{ij}=0 \label{eq:Bfree}\ed
\ee
The coefficient $\beta_{ij}$ represents a conserved $2$-form current, but instead of being a subleading mode it is actually diverging. Clearly, this does not respect the symmetries of AdS$_4$. It is nevertheless instructive to understand how the characteristic $2$-group Ward identity for $j\one$ emerges in this simple holographic model.

In order to see that, we need to compute the renormalized on-shell action for  \eqref{eq:holdualaction}. As before, the interaction does not affect the asymptotic behavior \eqref{eq:Bfree} but imposes the following relation 
\be
\partial_i\beta_i =\frac{ k}{8\pi^3} \beta_{ij}h_{ij}\ed \label{eq:dualcoeffrel}
\ee
Again, we will not provide the details of the full calculation and just present the resulting boundary variation
\be
\delta S_{\textrm{dual}}[b\two, a\one]+\delta S_{\textrm{c.t.}}[b\two, a\one]\Big|_{\textrm{on-shell}}=\int_{\textrm{bdry}} \delta\alpha_i \left(\beta_{i}-\frac{ k }{8\pi^3}\alpha_j\beta_{ij}\right)-\frac{1}{8\pi^2}\delta\alpha_{ij}\beta_{ij}\label{eq:dualonshell}\ec
\ee
from which we readily recognize the  currents
\begin{align}
j_i =\beta_{i}-\frac{ k}{8\pi^3}\alpha_j\beta_{ij}\ , \qquad 
J_{ij}=-\frac{1}{8\pi^2}\beta_{ij}\ed
\end{align}
Finally, making use of \eqref{eq:dualcoeffrel}, one gets the corresponding (non-)conservation equation
\be
\partial_i j_i =-\frac{ k}{2\pi}h_{ij}J_{ij}\ed
\label{eq:holdualward}
\ee
Note the appearance of the conserved current $J_{ij}$ on the right hand side of \eqref{eq:holdualward}, which is the standard indicator of a continuous $2$-group Ward identity. Equivalently, gauge invariance of \eqref{eq:dualonshell} is attained by the modified gauge transformations
\be
\alpha_i\to \alpha_i+\partial_i \lambda \ec \quad\alpha_{ij}\to \alpha_{ij}+\partial_{[i}\Lambda_{j]} -\frac{ k}{2\pi} \lambda h_{ij}\ec
\ee
thus consistently reproducing the 2-group gauge transformations. 

The present bottom-up holographic model reproduces the main features of a boundary three-dimensional quantum field theory exhibiting a $2$-group global symmetry. However this boundary theory is not conformal. Indeed, in getting to the renormalized action \eqref{eq:dualonshell} one has to add a counter-term which is roughly $R_{c}\beta_{ij}^2$, with $R_c$ the cut-off radius. The latter term is a double trace operator, and nothing else than the Maxwell kinetic term which in three dimensions is irrelevant and thus not scale invariant. 

Finally, let us briefly comment on a minimal holographic model realizing the non-invertible symmetries described in section \ref{sec:non-inv} in a dual boundary theory. We aim to describe a theory hosting the non-conservation equation on the right of \eqref{nonconv}. In order to achieve this in AdS$_4$, we need two $U(1)$ gauge fields $a\one$, $c\one$ together with a compact scalar $\chi$ with their dynamics described by the following action
\be
S_{\textrm{bulk}}[a\one,c\one, \chi] = \int   - \frac{1}{2}d\chi \wedge* d\chi -\frac{1}{2} f\two_a\wedge* f\two_a -\frac{1}{2} f\two_c\wedge* f\two_c +\frac{k}{4\pi^2} d\chi \wedge a\one \wedge f\two_c\ed
\label{eq:NIholoaction}
\ee
By solving the variational problem with Dirichlet boundary conditions for all the fields we get the following (non-)conservation equations for the boundary currents
\be
\partial_i j^A_i=-\frac{k}{4\pi^2}\epsilon^{ijk}\partial_i\alpha \partial_j\tilde\alpha_k \ec\qquad\partial_i j^C_i=0\ec
\ee
where $\alpha$, $\tilde\alpha_k$ are the leading coefficients in the asymptotic expansion of $\chi$ and $c\one$ respectively. At the boundary we can gauge  $U(1)\zero_C$, since its current is conserved, together with the periodic scalar background field whose symmetry is realized by $\chi$ in the bulk. Hence, upon switching the boundary conditions of $c\one$ and $\chi$ from Dirichlet to Neumann, we end up with the desired holographic description of the anomaly \eqref{nonconv}. Equivalently, we may dualize both fields by introducing new $1$-form and $2$-form gauge fields $\bar c\one$ and $b\two$ with Dirichlet boundary conditions such that
\be
d\bar c\one = *f_c\two-\frac{k}{4\pi^2}d\chi\wedge a\one\ec\qquad
d b\two = *d\chi- \frac{k}{4\pi^2} a\one\wedge f_c\two \ec
\label{ABJdual}
\ee
Let us mention that solving for the dual fields leads to a non-linear problem which in principle can be addressed by the methods described in \cite{Das:2022auy}, however we will not investigate this further in this work.

According to the analysis of section \ref{sec:non-inv}, the boundary theory hosts non-invertible symmetry defects for generic values of $k \in \bZ$. An important question in this context is how these operators are realized in the bulk. In this description one can interpret the boundary non-invertible symmetry defects $\cD$ also as defects of a higher codimension living in the bulk. Note that similar comments can be applied to the holographic model of $2$-group global symmetry presented before. This subject will be explored further in an upcoming work \cite{JADRAEGC}.

\bigskip
\begin{center}
\textbf{Acknowledgments}
\end{center}
We thank Eduardo Garcia-Valdecasas for collaboration at the initial stages of this work. We also thank Pietro Benetti Genolini, Francesco Benini, Daniel Brennan, Christian Copetti and Pierluigi Niro for helpful discussions. LT is grateful
for the hospitality of the King's College and SISSA where part of this work has been completed. JAD and RA are respectively a Postdoctoral Researcher and a Research Director of the F.R.S.-FNRS (Belgium). LT has been partially supported by funds from the Solvay Family. This research is further supported by IISN-Belgium (convention 4.4503.15) and by the F.R.S.-FNRS under the ``Excellence of Science" EOS be.h project n.~30820817.

\appendix
\section{Extra Examples}\label{apx}

\subsection{Compact Mass Parameters and Three-Dimensional Dualities}

In this section we describe a procedure to obtain a $2$-group global symmetry by considering a family of theories parametrized by a function of their coupling constants. Possibly, the simplest of such examples of 2-group symmetry arises already in the context of a single free Dirac fermion in three dimensions
\be
S[A\one, \psi]=\int i\bar \psi \slashed{D}_A \psi\ec \label{eq:freefermion}
\ee
where $A\one$ is a background for the $U(1)^{(0)}_A$ global symmetry. The free fermion theory can be gapped by deforming it with a mass term of the form $m\bar{\psi}\psi$. 

When $A\one=0$, the theory is trivially gapped for $m\to\pm\infty$. As a function of the mass parameter, the partition function $Z[m]$ then satisfies 
\be
\lim_{m\to \infty }Z[m]=\lim_{m\to -\infty}Z[m]=1\ed
\ee
Since both asymptotic limits are described by the same physics we can say that the mass parameter is effectively compact and parametrizes a circle, $m\in S^1$. 

However, when the $U(1)_A^{(0)}$ background is non-trivial, depending on the sign of $m$ the partition function of the gapped theory takes the following form
\be\label{contact3d}
\begin{split}
\lim_{m\to +\infty }Z[A\one, m]&=1\ec \\
\lim_{m\to -\infty }Z[A\one, m]&= \exp\left(-i\int_{\cY_4} \frac{1}{4\pi} dA\one\wedge dA\one + \frac{1}{96\pi} \Tr\left(R\two\wedge R\two\right)\right)\ed
\end{split}
\ee
where $\cY_4$ is a bulk extension of the spacetime three-manifold $\partial\cY_4 =\cM_3$ and $R\two$ is its Riemann curvature $2$-form. We refer the reader to \cite{Seiberg:2016gmd} for a careful explanation of why a gravitational Chern--Simons term appears on the right hand side of \eqref{contact3d}.

Following \cite{Seiberg:2018ntt,Cordova:2019jnf}, such non-invariance under the periodicity of $m$ can be interpreted as an anomaly in the space of coupling constants. In order to restore periodicity one should modify the partition function of the free fermion theory by introducing the following unusual counterterm 
\be
\tilde Z[A\one, m]=Z[A\one, m]\exp\left(-i\int_{\cY_4} \frac{1}{8\pi^2}\rho(m) dA\one\wedge dA\one + \frac{1}{192\pi^2}\rho(m)  \Tr\left(R\two\wedge R\two\right)\right)\ec \label{eq:tildeZ}
\ee
with a function $\rho(m)$ such that $\lim_{m\to \infty}(\rho(m)-\rho(-m))= 2\pi$. The function $\rho(m)$ has non-trivial winding in the mass parameter space which is taken to be a circle $S^1$. Following our discussion from the main text, we interpret it here as a background compact scalar field. In particular, we can write the counterterm in \eqref{eq:tildeZ} as a five-dimensional anomaly inflow term of the form
\be
{\cal I}_5 = -\int \frac{1}{8\pi^2}d\rho\wedge dA\one\wedge dA\one + \frac{1}{192\pi^2}d\rho \wedge\Tr\left(R\two\wedge R\two\right)\ed\label{eq:I5rho}
\ee
The structure of \eqref{eq:I5rho} is similar to the type of anomalies described in the main text. This suggests that we should be able to promote $\rho$ to a dynamical field and activate a background field for its associated $U(1)^{(1)}_B$ winding symmetry. The theory obtained after gauging $\rho$ would then lead to a continuous $2$-group global symmetry,
\be\label{3dfer2grp}
(U(1)^{(0)}_A\times \mathscr{P})\times_{\kappa_A,\kappa_{\mathscr{P}}} U(1)^{(1)}_B\ec
\ee
where $\mathscr{P}$ denotes the Poincaré spacetime symmetry.

The theory \eqref{eq:freefermion} features in the simplest three-dimensional bosonization duality \cite{Seiberg:2016gmd,Polyakov:1988md,Aharony:2015mjs,Karch:2016sxi}
\be\label{3dduality}
i\bar\psi\slashed{D}_A\psi \longleftrightarrow |D_{b}\phi|-|\phi|^4+\frac{1}{4\pi}b\one db\one+\frac{1}{2\pi}b\one dA_T\one\ec
\ee
that is, a duality between a free fermion and a gauged version of the $O(2)$ Wilson--Fisher fixed point. In the above duality, the $U(1)_A^{(0)}$ flavor symmetry on the fermionic side is mapped to the $U(1)_{A_T}\zero$ topological symmetry acting on monopole operators in the bosonic description.

The massive phases on both sides of the duality are also matched. One can interpret the fermionic mass operator $m\bar\psi\psi$ as the operator $-m^2|\phi|^2$ in the bosonic theory. Then, for $m\to\infty$, $\phi$ condenses and completely Higgses the $U(1)_b$ gauge symmetry. In this limit the theory is trivially gapped. On the other end, for $m\to-\infty$, $\phi$ is massive and decouples but $U(1)_b$ remains. In this phase, the massive boson $\phi$ statistics is modified by the flux attachment phenomenon \cite{Wilczek:1981du,Jain:1989tx} due to the $U(1)_1$ Chern--Simons coupling
\be
\begin{split}
\lim_{m\to +\infty }Z[A_T\one, m]&=1\ec\\
\lim_{m\to -\infty }Z[A_T\one, m]&= \exp\left(-i\int \frac{1}{4\pi}dA_T\one\wedge dA_T\one + \frac{1}{96\pi}\Tr\left(R\two \wedge R\two\right) \right)\ec
\end{split}
\ee
see again \cite{Seiberg:2016gmd} for details on the gravitational term.
In absence of background fields, we may identify both mass asymptotic limits and take $m\in S^1$. Therefore also the Wilson--Fisher boson has a mixed anomaly \eqref{eq:I5rho}. As a result, upon gauging $\rho$ on the bosonic side we obtain a $2$-group global symmetry that perfectly matches the one found on the fermionic side \eqref{3dfer2grp} by the obvious identification $A\one=A\one_T$.\footnote{A map of $2$-group global symmetries across dualities have been recently investigated for certain Seiberg dual pairs of four-dimensional supersymmetric gauge theories in \cite{Lee:2021crt}.}

Before concluding this subsection we would like to point out a subtle issue regarding the quantization of the coefficients $\kappa_A, \kappa_{\mathscr{P}}$ appearing in \eqref{3dfer2grp}. In four dimensions (see section 7.1 of \cite{CDI}) the relevant background $2$-form field $B\two$ transforms as follows
\be
B\two \to B\two +d\Lambda\one_B + \frac{\kappa_A}{4\pi}\lambda\zero_A F\two_A + \frac{\kappa_{\mathscr{P}}}{96\pi}\Tr(\lambda\zero_\omega d\omega\one)\ec
\ee
where $\lambda\zero_\omega$ is a local frame rotation and $\omega\one$ is a spin connection. Such a $2$-group transformation is free of ambiguity from large gauge transformations of $\lambda\zero_A$ and $\lambda\zero_\omega$ if and only if
\be\label{quant4d}
\kappa_A \in 2\bZ\ec\qquad \kappa_{\mathscr{P}}\in 6\bZ\ed
\ee
Furthermore, these quantization conditions can be determined by analyzing the integrality properties of the six-forms that enter the anomaly polynomial using the Atiyah-Singer index theorem. It is not clear to us if $\kappa_A$ and $\kappa_{\mathscr{P}}$ should still obey the same quantization conditions in three dimensions. In particular, it appears that the explicit values $\kappa_A=\kappa_{\mathscr{P}}=1$ that we can extract from the anomaly \eqref{eq:I5rho} would be in tension with \eqref{quant4d}.

\subsection{KK Reductions}

A related example is the following. It can be defined from the reduction of a theory in four dimensions to one in three dimensions, where one can define a $2$-group only in the reduced theory. For instance in four dimensions one can have a single $U(1)_A\zero$ symmetry with a cubic 't Hooft anomaly $k_{A^3}$. Upon reduction to three dimensions, one can contemplate gauging only the compact scalar, while leaving the vector to its background status. We can then have a genuine $2$-group in three dimensions that has no origin in four dimensions.

For the simplest example, consider the reduction of a single  Weyl fermion in four dimensions to a KK tower of Dirac fermions in three dimensions. The coupling to the component $\theta$ of the background gauge field for the chiral $U(1)\zero_A$ symmetry along the KK circle manifests itself as a $\theta$-dependent correction to the KK masses in three dimensions
\begin{align}
    m_n=m_0+\frac{2\pi n}{L}= \frac{1}{L}(\theta+2\pi n)\ .
\end{align}
It is obvious that $\theta$ has periodicity $\theta \sim \theta + 2\pi$ since that operation does not change the spectrum of masses of the infinite tower. This periodicity is consistent with the presence of a single $U(1)\zero_A$ which rotates all the fermions of the tower simultaneously.

However, the $U(1)\zero_A$ has a mixed anomaly with the periodicity of $\theta$. It descends straightforwardly from the cubic anomaly a single Weyl fermion in four dimensions, and yields the following anomaly polynomial (for simplicity we neglect the gravitational part of the anomaly)
\be
{\cal I}_5 = \int \frac{1}{8\pi^2}d\rho\wedge dA\one\wedge dA\one\ed
\ee
In order to see the anomaly arise purely in three dimensions, one can turn on a background gauge field $A\one$ for the $U(1)\zero_A$ and compute its effective action upon integrating over the whole tower of fermions. This is done in a related model in \cite{Ohmori:2021sqg,DiPietro:2014bca}. We can then gauge the shift symmetry $\theta\sim \theta +2\pi $. The scalar has now a topological $U(1)\one_B$ $1$-form symmetry whose current is $J\two_B=\star d\phi/2\pi$, which participates in a $2$-group with the $0$-form symmetry $U(1)\zero_A$. 

Note that this model displays the same subtlety as those in the previous subsection. Indeed the transformation of the background $B\two$ field would still be realized with $\kappa_A={k_A^3}=1$, thus violating \eqref{quant4d}. The problem cannot be linked to the fact that we are partially gauging a symmetry that has a cubic anomaly in four dimensions. Indeed, one can just take two Weyl fermions with the same unit charge in four dimensions, so that the cubic anomaly is still non-zero. Upon performing all the steps above, one finds that now $\kappa_A=k_{A^3}=2$, in agreement with \eqref{quant4d}.

\makeatletter
\interlinepenalty=10000
\bibliographystyle{utphys}
\bibliography{GenSymBib}

\providecommand{\href}[2]{#2}\begingroup\raggedright\begin{thebibliography}{10}

\bibitem{GKSW}
D.~Gaiotto, A.~Kapustin, N.~Seiberg, and B.~Willett, ``{Generalized Global
  Symmetries},'' \href{http://dx.doi.org/10.1007/JHEP02(2015)172}{{\em JHEP}
  {\bfseries 02} (2015) 172}, \href{http://arxiv.org/abs/1412.5148}{{\ttfamily
  arXiv:1412.5148 [hep-th]}}.

\bibitem{Kapustin:2013uxa}
A.~Kapustin and R.~Thorngren, ``{Higher symmetry and gapped phases of gauge
  theories},'' \href{http://arxiv.org/abs/1309.4721}{{\ttfamily arXiv:1309.4721
  [hep-th]}}.

\bibitem{Sharpe:2015mja}
E.~Sharpe, ``{Notes on generalized global symmetries in QFT},''
  \href{http://dx.doi.org/10.1002/prop.201500048}{{\em Fortsch. Phys.}
  {\bfseries 63} (2015) 659--682},
  \href{http://arxiv.org/abs/1508.04770}{{\ttfamily arXiv:1508.04770
  [hep-th]}}.

\bibitem{Thorngren:2015gtw}
R.~Thorngren and C.~von Keyserlingk, ``{Higher SPT's and a generalization of
  anomaly in-flow},'' \href{http://arxiv.org/abs/1511.02929}{{\ttfamily
  arXiv:1511.02929 [cond-mat.str-el]}}.

\bibitem{Tachikawa:2017gyf}
Y.~Tachikawa, ``{On gauging finite subgroups},''
  \href{http://dx.doi.org/10.21468/SciPostPhys.8.1.015}{{\em SciPost Phys.}
  {\bfseries 8} no.~1, (2020) 015},
  \href{http://arxiv.org/abs/1712.09542}{{\ttfamily arXiv:1712.09542
  [hep-th]}}.

\bibitem{CDI}
C.~C\'ordova, T.~T. Dumitrescu, and K.~Intriligator, ``{Exploring 2-Group
  Global Symmetries},'' \href{http://dx.doi.org/10.1007/JHEP02(2019)184}{{\em
  JHEP} {\bfseries 02} (2019) 184},
  \href{http://arxiv.org/abs/1802.04790}{{\ttfamily arXiv:1802.04790
  [hep-th]}}.

\bibitem{Delcamp:2018wlb}
C.~Delcamp and A.~Tiwari, ``{From gauge to higher gauge models of topological
  phases},'' \href{http://dx.doi.org/10.1007/JHEP10(2018)049}{{\em JHEP}
  {\bfseries 10} (2018) 049}, \href{http://arxiv.org/abs/1802.10104}{{\ttfamily
  arXiv:1802.10104 [cond-mat.str-el]}}.

\bibitem{Benini:2018reh}
F.~Benini, C.~C\'ordova, and P.-S. Hsin, ``{On 2-Group Global Symmetries and
  their Anomalies},'' \href{http://dx.doi.org/10.1007/JHEP03(2019)118}{{\em
  JHEP} {\bfseries 03} (2019) 118},
  \href{http://arxiv.org/abs/1803.09336}{{\ttfamily arXiv:1803.09336
  [hep-th]}}.

\bibitem{BaezLauda1}
J.~C. {Baez} and A.~D. {Lauda}, ``{Higher-Dimensional Algebra V: 2-Groups},''
  \href{http://arxiv.org/abs/math/0307200}{{\ttfamily arXiv:math/0307200
  [math.QA]}}.

\bibitem{Baez:2004in}
J.~Baez and U.~Schreiber, ``{Higher gauge theory: 2-connections on
  2-bundles},'' \href{http://arxiv.org/abs/hep-th/0412325}{{\ttfamily
  arXiv:hep-th/0412325}}.

\bibitem{SchreiberWaldorf}
U.~{Schreiber} and K.~{Waldorf}, ``{Connections on non-abelian Gerbes and their
  Holonomy},'' \href{http://arxiv.org/abs/arXiv:0808.1923}{{\ttfamily
  arXiv:arXiv:0808.1923}}.

\bibitem{Hsin:2020nts}
P.-S. Hsin and H.~T. Lam, ``{Discrete theta angles, symmetries and
  anomalies},'' \href{http://dx.doi.org/10.21468/SciPostPhys.10.2.032}{{\em
  SciPost Phys.} {\bfseries 10} no.~2, (2021) 032},
  \href{http://arxiv.org/abs/2007.05915}{{\ttfamily arXiv:2007.05915
  [hep-th]}}.

\bibitem{Gukov:2020btk}
S.~Gukov, P.-S. Hsin, and D.~Pei, ``{Generalized global symmetries of $T[M]$
  theories. Part I},'' \href{http://dx.doi.org/10.1007/JHEP04(2021)232}{{\em
  JHEP} {\bfseries 04} (2021) 232},
  \href{http://arxiv.org/abs/2010.15890}{{\ttfamily arXiv:2010.15890
  [hep-th]}}.

\bibitem{Cordova:2020tij}
C.~Cordova, T.~T. Dumitrescu, and K.~Intriligator, ``{2-Group Global Symmetries
  and Anomalies in Six-Dimensional Quantum Field Theories},''
  \href{http://dx.doi.org/10.1007/JHEP04(2021)252}{{\em JHEP} {\bfseries 04}
  (2021) 252}, \href{http://arxiv.org/abs/2009.00138}{{\ttfamily
  arXiv:2009.00138 [hep-th]}}.

\bibitem{DelZotto:2020sop}
M.~Del~Zotto and K.~Ohmori, ``{2-Group Symmetries of 6D Little String Theories
  and T-Duality},'' \href{http://dx.doi.org/10.1007/s00023-021-01018-3}{{\em
  Annales Henri Poincare} {\bfseries 22} no.~7, (2021) 2451--2474},
  \href{http://arxiv.org/abs/2009.03489}{{\ttfamily arXiv:2009.03489
  [hep-th]}}.

\bibitem{Apruzzi:2021vcu}
F.~Apruzzi, L.~Bhardwaj, J.~Oh, and S.~Schafer-Nameki, ``{The Global Form of
  Flavor Symmetries and 2-Group Symmetries in 5d SCFTs},''
  \href{http://arxiv.org/abs/2105.08724}{{\ttfamily arXiv:2105.08724
  [hep-th]}}.

\bibitem{Lee:2021crt}
Y.~Lee, K.~Ohmori, and Y.~Tachikawa, ``{Matching higher symmetries across
  Intriligator-Seiberg duality},''
  \href{http://dx.doi.org/10.1007/JHEP10(2021)114}{{\em JHEP} {\bfseries 10}
  (2021) 114}, \href{http://arxiv.org/abs/2108.05369}{{\ttfamily
  arXiv:2108.05369 [hep-th]}}.

\bibitem{Apruzzi:2021mlh}
F.~Apruzzi, L.~Bhardwaj, D.~S.~W. Gould, and S.~Schafer-Nameki, ``{2-Group
  symmetries and their classification in 6d},''
  \href{http://dx.doi.org/10.21468/SciPostPhys.12.3.098}{{\em SciPost Phys.}
  {\bfseries 12} no.~3, (2022) 098},
  \href{http://arxiv.org/abs/2110.14647}{{\ttfamily arXiv:2110.14647
  [hep-th]}}.

\bibitem{Delacretaz:2019brr}
L.~V. Delacr\'etaz, D.~M. Hofman, and G.~Mathys, ``{Superfluids as Higher-form
  Anomalies},'' \href{http://dx.doi.org/10.21468/SciPostPhys.8.3.047}{{\em
  SciPost Phys.} {\bfseries 8} (2020) 047},
  \href{http://arxiv.org/abs/1908.06977}{{\ttfamily arXiv:1908.06977
  [hep-th]}}.

\bibitem{Seiberg:2018ntt}
N.~Seiberg, Y.~Tachikawa, and K.~Yonekura, ``{Anomalies of Duality Groups and
  Extended Conformal Manifolds},''
  \href{http://dx.doi.org/10.1093/ptep/pty069}{{\em PTEP} {\bfseries 2018}
  no.~7, (2018) 073B04}, \href{http://arxiv.org/abs/1803.07366}{{\ttfamily
  arXiv:1803.07366 [hep-th]}}.

\bibitem{Cordova:2019jnf}
C.~C\'ordova, D.~S. Freed, H.~T. Lam, and N.~Seiberg, ``{Anomalies in the Space
  of Coupling Constants and Their Dynamical Applications I},''
  \href{http://dx.doi.org/10.21468/SciPostPhys.8.1.001}{{\em SciPost Phys.}
  {\bfseries 8} no.~1, (2020) 001},
  \href{http://arxiv.org/abs/1905.09315}{{\ttfamily arXiv:1905.09315
  [hep-th]}}.

\bibitem{Cordova:2019uob}
C.~C\'ordova, D.~S. Freed, H.~T. Lam, and N.~Seiberg, ``{Anomalies in the Space
  of Coupling Constants and Their Dynamical Applications II},''
  \href{http://dx.doi.org/10.21468/SciPostPhys.8.1.002}{{\em SciPost Phys.}
  {\bfseries 8} no.~1, (2020) 002},
  \href{http://arxiv.org/abs/1905.13361}{{\ttfamily arXiv:1905.13361
  [hep-th]}}.

\bibitem{Hsin:2020cgg}
P.-S. Hsin, A.~Kapustin, and R.~Thorngren, ``{Berry Phase in Quantum Field
  Theory: Diabolical Points and Boundary Phenomena},''
  \href{http://dx.doi.org/10.1103/PhysRevB.102.245113}{{\em Phys. Rev. B}
  {\bfseries 102} (2020) 245113},
  \href{http://arxiv.org/abs/2004.10758}{{\ttfamily arXiv:2004.10758
  [cond-mat.str-el]}}.

\bibitem{Kapustin:2020eby}
A.~Kapustin and L.~Spodyneiko, ``{Higher-dimensional generalizations of Berry
  curvature},'' \href{http://dx.doi.org/10.1103/PhysRevB.101.235130}{{\em Phys.
  Rev. B} {\bfseries 101} no.~23, (2020) 235130},
  \href{http://arxiv.org/abs/2001.03454}{{\ttfamily arXiv:2001.03454
  [cond-mat.str-el]}}.

\bibitem{Choi:2022odr}
Y.~Choi and K.~Ohmori, ``{Higher Berry Phase of Fermions and Index Theorem},''
  \href{http://arxiv.org/abs/2205.02188}{{\ttfamily arXiv:2205.02188
  [hep-th]}}.

\bibitem{Benini:2017dus}
F.~Benini, P.-S. Hsin, and N.~Seiberg, ``{Comments on global symmetries,
  anomalies, and duality in (2 + 1)d},''
  \href{http://dx.doi.org/10.1007/JHEP04(2017)135}{{\em JHEP} {\bfseries 04}
  (2017) 135}, \href{http://arxiv.org/abs/1702.07035}{{\ttfamily
  arXiv:1702.07035 [cond-mat.str-el]}}.

\bibitem{Genolini:2022mpi}
P.~B. Genolini and L.~Tizzano, ``{Comments on Global Symmetries and Anomalies
  of $5d$ SCFTs},'' \href{http://arxiv.org/abs/2201.02190}{{\ttfamily
  arXiv:2201.02190 [hep-th]}}.

\bibitem{Bhardwaj:2022dyt}
L.~Bhardwaj, M.~Bullimore, A.~E.~V. Ferrari, and S.~Schafer-Nameki,
  ``{Anomalies of Generalized Symmetries from Solitonic Defects},''
  \href{http://arxiv.org/abs/2205.15330}{{\ttfamily arXiv:2205.15330
  [hep-th]}}.

\bibitem{Gaiotto:2017yup}
D.~Gaiotto, A.~Kapustin, Z.~Komargodski, and N.~Seiberg, ``{Theta, Time
  Reversal, and Temperature},''
  \href{http://dx.doi.org/10.1007/JHEP05(2017)091}{{\em JHEP} {\bfseries 05}
  (2017) 091}, \href{http://arxiv.org/abs/1703.00501}{{\ttfamily
  arXiv:1703.00501 [hep-th]}}.

\bibitem{Choi:2022jqy}
Y.~Choi, H.~T. Lam, and S.-H. Shao, ``{Non-invertible Global Symmetries in the
  Standard Model},'' \href{http://arxiv.org/abs/2205.05086}{{\ttfamily
  arXiv:2205.05086 [hep-th]}}.

\bibitem{Cordova:2022ieu}
C.~Cordova and K.~Ohmori, ``{Non-Invertible Chiral Symmetry and Exponential
  Hierarchies},'' \href{http://arxiv.org/abs/2205.06243}{{\ttfamily
  arXiv:2205.06243 [hep-th]}}.

\bibitem{Frohlich:2006ch}
J.~Frohlich, J.~Fuchs, I.~Runkel, and C.~Schweigert, ``{Duality and defects in
  rational conformal field theory},''
  \href{http://dx.doi.org/10.1016/j.nuclphysb.2006.11.017}{{\em Nucl. Phys. B}
  {\bfseries 763} (2007) 354--430},
  \href{http://arxiv.org/abs/hep-th/0607247}{{\ttfamily arXiv:hep-th/0607247}}.

\bibitem{Bhardwaj:2017xup}
L.~Bhardwaj and Y.~Tachikawa, ``{On finite symmetries and their gauging in two
  dimensions},'' \href{http://dx.doi.org/10.1007/JHEP03(2018)189}{{\em JHEP}
  {\bfseries 03} (2018) 189}, \href{http://arxiv.org/abs/1704.02330}{{\ttfamily
  arXiv:1704.02330 [hep-th]}}.

\bibitem{Chang:2018iay}
C.-M. Chang, Y.-H. Lin, S.-H. Shao, Y.~Wang, and X.~Yin, ``{Topological Defect
  Lines and Renormalization Group Flows in Two Dimensions},''
  \href{http://dx.doi.org/10.1007/JHEP01(2019)026}{{\em JHEP} {\bfseries 01}
  (2019) 026}, \href{http://arxiv.org/abs/1802.04445}{{\ttfamily
  arXiv:1802.04445 [hep-th]}}.

\bibitem{Thorngren:2019iar}
R.~Thorngren and Y.~Wang, ``{Fusion Category Symmetry I: Anomaly In-Flow and
  Gapped Phases},'' \href{http://arxiv.org/abs/1912.02817}{{\ttfamily
  arXiv:1912.02817 [hep-th]}}.

\bibitem{Gaiotto:2020iye}
D.~Gaiotto and J.~Kulp, ``{Orbifold groupoids},''
  \href{http://dx.doi.org/10.1007/JHEP02(2021)132}{{\em JHEP} {\bfseries 02}
  (2021) 132}, \href{http://arxiv.org/abs/2008.05960}{{\ttfamily
  arXiv:2008.05960 [hep-th]}}.

\bibitem{Komargodski:2020mxz}
Z.~Komargodski, K.~Ohmori, K.~Roumpedakis, and S.~Seifnashri, ``{Symmetries and
  strings of adjoint QCD$_{2}$},''
  \href{http://dx.doi.org/10.1007/JHEP03(2021)103}{{\em JHEP} {\bfseries 03}
  (2021) 103}, \href{http://arxiv.org/abs/2008.07567}{{\ttfamily
  arXiv:2008.07567 [hep-th]}}.

\bibitem{Nguyen:2021yld}
M.~Nguyen, Y.~Tanizaki, and M.~\"Unsal, ``{Semi-Abelian gauge theories,
  non-invertible symmetries, and string tensions beyond $N$-ality},''
  \href{http://dx.doi.org/10.1007/JHEP03(2021)238}{{\em JHEP} {\bfseries 03}
  (2021) 238}, \href{http://arxiv.org/abs/2101.02227}{{\ttfamily
  arXiv:2101.02227 [hep-th]}}.

\bibitem{Thorngren:2021yso}
R.~Thorngren and Y.~Wang, ``{Fusion Category Symmetry II: Categoriosities at
  $c$ = 1 and Beyond},'' \href{http://arxiv.org/abs/2106.12577}{{\ttfamily
  arXiv:2106.12577 [hep-th]}}.

\bibitem{Koide:2021zxj}
M.~Koide, Y.~Nagoya, and S.~Yamaguchi, ``{Non-invertible topological defects in
  4-dimensional $\mathbb {Z}_2$ pure lattice gauge theory},''
  \href{http://dx.doi.org/10.1093/ptep/ptab145}{{\em PTEP} {\bfseries 2022}
  no.~1, (2022) 013B03}, \href{http://arxiv.org/abs/2109.05992}{{\ttfamily
  arXiv:2109.05992 [hep-th]}}.

\bibitem{Choi:2021kmx}
Y.~Choi, C.~Cordova, P.-S. Hsin, H.~T. Lam, and S.-H. Shao, ``{Non-Invertible
  Duality Defects in 3+1 Dimensions},''
  \href{http://arxiv.org/abs/2111.01139}{{\ttfamily arXiv:2111.01139
  [hep-th]}}.

\bibitem{Kaidi:2021xfk}
J.~Kaidi, K.~Ohmori, and Y.~Zheng, ``{Kramers-Wannier-like Duality Defects in
  (3+1)D Gauge Theories},''
  \href{http://dx.doi.org/10.1103/PhysRevLett.128.111601}{{\em Phys. Rev.
  Lett.} {\bfseries 128} no.~11, (2022) 111601},
  \href{http://arxiv.org/abs/2111.01141}{{\ttfamily arXiv:2111.01141
  [hep-th]}}.

\bibitem{Burbano:2021loy}
I.~M. Burbano, J.~Kulp, and J.~Neuser, ``{Duality defects in E$_{8}$},''
  \href{http://dx.doi.org/10.1007/JHEP10(2022)187}{{\em JHEP} {\bfseries 10}
  (2022) 186}, \href{http://arxiv.org/abs/2112.14323}{{\ttfamily
  arXiv:2112.14323 [hep-th]}}.

\bibitem{Choi:2022zal}
Y.~Choi, C.~Cordova, P.-S. Hsin, H.~T. Lam, and S.-H. Shao, ``{Non-invertible
  Condensation, Duality, and Triality Defects in 3+1 Dimensions},''
  \href{http://arxiv.org/abs/2204.09025}{{\ttfamily arXiv:2204.09025
  [hep-th]}}.

\bibitem{Roumpedakis:2022aik}
K.~Roumpedakis, S.~Seifnashri, and S.-H. Shao, ``{Higher Gauging and
  Non-invertible Condensation Defects},''
  \href{http://arxiv.org/abs/2204.02407}{{\ttfamily arXiv:2204.02407
  [hep-th]}}.

\bibitem{Hayashi:2022fkw}
Y.~Hayashi and Y.~Tanizaki, ``{Non-invertible self-duality defects of
  Cardy-Rabinovici model and mixed gravitational anomaly},''
  \href{http://arxiv.org/abs/2204.07440}{{\ttfamily arXiv:2204.07440
  [hep-th]}}.

\bibitem{Arias-Tamargo:2022nlf}
G.~Arias-Tamargo and D.~Rodriguez-Gomez, ``{Non-Invertible Symmetries from
  Discrete Gauging and Completeness of the Spectrum},''
  \href{http://arxiv.org/abs/2204.07523}{{\ttfamily arXiv:2204.07523
  [hep-th]}}.

\bibitem{Bhardwaj:2022yxj}
L.~Bhardwaj, L.~Bottini, S.~Schafer-Nameki, and A.~Tiwari, ``{Non-Invertible
  Higher-Categorical Symmetries},''
  \href{http://arxiv.org/abs/2204.06564}{{\ttfamily arXiv:2204.06564
  [hep-th]}}.

\bibitem{Kaidi:2022uux}
J.~Kaidi, G.~Zafrir, and Y.~Zheng, ``{Non-Invertible Symmetries of
  $\mathcal{N}=4$ SYM and Twisted Compactification},''
  \href{http://arxiv.org/abs/2205.01104}{{\ttfamily arXiv:2205.01104
  [hep-th]}}.

\bibitem{Antinucci:2022eat}
A.~Antinucci, G.~Galati, and G.~Rizi, ``{On Continuous 2-Category Symmetries
  and Yang-Mills Theory},'' \href{http://arxiv.org/abs/2206.05646}{{\ttfamily
  arXiv:2206.05646 [hep-th]}}.

\bibitem{Bashmakov:2022jtl}
V.~Bashmakov, M.~Del~Zotto, and A.~Hasan, ``{On the 6d Origin of Non-invertible
  Symmetries in 4d},'' \href{http://arxiv.org/abs/2206.07073}{{\ttfamily
  arXiv:2206.07073 [hep-th]}}.

\bibitem{Hofman:2018lfz}
D.~M. Hofman and N.~Iqbal, ``{Goldstone modes and photonization for higher form
  symmetries},'' \href{http://dx.doi.org/10.21468/SciPostPhys.6.1.006}{{\em
  SciPost Phys.} {\bfseries 6} no.~1, (2019) 006},
  \href{http://arxiv.org/abs/1802.09512}{{\ttfamily arXiv:1802.09512
  [hep-th]}}.

\bibitem{Lake:2018dqm}
E.~Lake, ``{Higher-form symmetries and spontaneous symmetry breaking},''
  \href{http://arxiv.org/abs/1802.07747}{{\ttfamily arXiv:1802.07747
  [hep-th]}}.

\bibitem{Brauner:2020rtz}
T.~Brauner, ``{Field theories with higher-group symmetry from composite
  currents},'' \href{http://dx.doi.org/10.1007/JHEP04(2021)045}{{\em JHEP}
  {\bfseries 04} (2021) 045}, \href{http://arxiv.org/abs/2012.00051}{{\ttfamily
  arXiv:2012.00051 [hep-th]}}.

\bibitem{Heidenreich:2020pkc}
B.~Heidenreich, J.~McNamara, M.~Montero, M.~Reece, T.~Rudelius, and
  I.~Valenzuela, ``{Chern-Weil global symmetries and how quantum gravity avoids
  them},'' \href{http://dx.doi.org/10.1007/JHEP11(2021)053}{{\em JHEP}
  {\bfseries 11} (2021) 053}, \href{http://arxiv.org/abs/2012.00009}{{\ttfamily
  arXiv:2012.00009 [hep-th]}}.

\bibitem{Witten:1979ey}
E.~Witten, ``{Dyons of Charge e theta/2 pi},''
  \href{http://dx.doi.org/10.1016/0370-2693(79)90838-4}{{\em Phys. Lett. B}
  {\bfseries 86} (1979) 283--287}.

\bibitem{Brennan:2020ehu}
T.~D. Brennan and C.~Cordova, ``{Axions, higher-groups, and emergent
  symmetry},'' \href{http://dx.doi.org/10.1007/JHEP02(2022)145}{{\em JHEP}
  {\bfseries 02} (2022) 145}, \href{http://arxiv.org/abs/2011.09600}{{\ttfamily
  arXiv:2011.09600 [hep-th]}}.

\bibitem{Hidaka:2020iaz}
Y.~Hidaka, M.~Nitta, and R.~Yokokura, ``{Higher-form symmetries and 3-group in
  axion electrodynamics},''
  \href{http://dx.doi.org/10.1016/j.physletb.2020.135672}{{\em Phys. Lett. B}
  {\bfseries 808} (2020) 135672},
  \href{http://arxiv.org/abs/2006.12532}{{\ttfamily arXiv:2006.12532
  [hep-th]}}.

\bibitem{Hidaka:2021mml}
Y.~Hidaka, M.~Nitta, and R.~Yokokura, ``{Topological axion electrodynamics and
  4-group symmetry},''
  \href{http://dx.doi.org/10.1016/j.physletb.2021.136762}{{\em Phys. Lett. B}
  {\bfseries 823} (2021) 136762},
  \href{http://arxiv.org/abs/2107.08753}{{\ttfamily arXiv:2107.08753
  [hep-th]}}.

\bibitem{BenettiGenolini:2020doj}
P.~Benetti~Genolini and L.~Tizzano, ``{Instantons, symmetries and anomalies in
  five dimensions},'' \href{http://dx.doi.org/10.1007/JHEP04(2021)188}{{\em
  JHEP} {\bfseries 04} (2021) 188},
  \href{http://arxiv.org/abs/2009.07873}{{\ttfamily arXiv:2009.07873
  [hep-th]}}.

\bibitem{Ohmori:2021sqg}
K.~Ohmori and L.~Tizzano, ``{Anomaly Matching Across Dimensions and
  Supersymmetric Cardy Formulae},''
  \href{http://arxiv.org/abs/2112.13445}{{\ttfamily arXiv:2112.13445
  [hep-th]}}.

\bibitem{DiPietro:2014bca}
L.~Di~Pietro and Z.~Komargodski, ``{Cardy formulae for SUSY theories in $d =$ 4
  and $d =$ 6},'' \href{http://dx.doi.org/10.1007/JHEP12(2014)031}{{\em JHEP}
  {\bfseries 12} (2014) 031}, \href{http://arxiv.org/abs/1407.6061}{{\ttfamily
  arXiv:1407.6061 [hep-th]}}.

\bibitem{Kapustin:2014gua}
A.~Kapustin and N.~Seiberg, ``{Coupling a QFT to a TQFT and Duality},''
  \href{http://dx.doi.org/10.1007/JHEP04(2014)001}{{\em JHEP} {\bfseries 04}
  (2014) 001}, \href{http://arxiv.org/abs/1401.0740}{{\ttfamily arXiv:1401.0740
  [hep-th]}}.

\bibitem{Hofman:2017vwr}
D.~M. Hofman and N.~Iqbal, ``{Generalized global symmetries and holography},''
  \href{http://dx.doi.org/10.21468/SciPostPhys.4.1.005}{{\em SciPost Phys.}
  {\bfseries 4} no.~1, (2018) 005},
  \href{http://arxiv.org/abs/1707.08577}{{\ttfamily arXiv:1707.08577
  [hep-th]}}.

\bibitem{Grozdanov:2017kyl}
S.~Grozdanov and N.~Poovuttikul, ``{Generalised global symmetries in
  holography: magnetohydrodynamic waves in a strongly interacting plasma},''
  \href{http://dx.doi.org/10.1007/JHEP04(2019)141}{{\em JHEP} {\bfseries 04}
  (2019) 141}, \href{http://arxiv.org/abs/1707.04182}{{\ttfamily
  arXiv:1707.04182 [hep-th]}}.

\bibitem{Iqbal:2020lrt}
N.~Iqbal and N.~Poovuttikul, ``{2-group global symmetries, hydrodynamics and
  holography},'' \href{http://arxiv.org/abs/2010.00320}{{\ttfamily
  arXiv:2010.00320 [hep-th]}}.

\bibitem{Iqbal:2021tui}
N.~Iqbal and K.~Macfarlane, ``{Higher-form symmetry breaking and holographic
  flavour},'' \href{http://arxiv.org/abs/2107.00373}{{\ttfamily
  arXiv:2107.00373 [hep-th]}}.

\bibitem{DeWolfe:2020uzb}
O.~DeWolfe and K.~Higginbotham, ``{Generalized symmetries and 2-groups via
  electromagnetic duality in $AdS/CFT$},''
  \href{http://dx.doi.org/10.1103/PhysRevD.103.026011}{{\em Phys. Rev. D}
  {\bfseries 103} no.~2, (2021) 026011},
  \href{http://arxiv.org/abs/2010.06594}{{\ttfamily arXiv:2010.06594
  [hep-th]}}.

\bibitem{Lee:2021obi}
Y.~Lee and Y.~Zheng, ``{Remarks on compatibility between conformal symmetry and
  continuous higher-form symmetries},''
  \href{http://dx.doi.org/10.1103/PhysRevD.104.085005}{{\em Phys. Rev. D}
  {\bfseries 104} no.~8, (2021) 085005},
  \href{http://arxiv.org/abs/2108.00732}{{\ttfamily arXiv:2108.00732
  [hep-th]}}.

\bibitem{Breitenlohner:1982bm}
P.~Breitenlohner and D.~Z. Freedman, ``{Positive Energy in anti-De Sitter
  Backgrounds and Gauged Extended Supergravity},''
  \href{http://dx.doi.org/10.1016/0370-2693(82)90643-8}{{\em Phys. Lett. B}
  {\bfseries 115} (1982) 197--201}.

\bibitem{Andrade:2011dg}
T.~Andrade and D.~Marolf, ``{AdS/CFT beyond the unitarity bound},''
  \href{http://dx.doi.org/10.1007/JHEP01(2012)049}{{\em JHEP} {\bfseries 01}
  (2012) 049}, \href{http://arxiv.org/abs/1105.6337}{{\ttfamily arXiv:1105.6337
  [hep-th]}}.

\bibitem{Witten:2003ya}
E.~Witten, ``{SL(2,Z) action on three-dimensional conformal field theories with
  Abelian symmetry},'' \href{http://arxiv.org/abs/hep-th/0307041}{{\ttfamily
  arXiv:hep-th/0307041}}.

\bibitem{Witten:1998wy}
E.~Witten, ``{AdS / CFT correspondence and topological field theory},''
  \href{http://dx.doi.org/10.1088/1126-6708/1998/12/012}{{\em JHEP} {\bfseries
  12} (1998) 012}, \href{http://arxiv.org/abs/hep-th/9812012}{{\ttfamily
  arXiv:hep-th/9812012}}.

\bibitem{Marolf:2006nd}
D.~Marolf and S.~F. Ross, ``{Boundary Conditions and New Dualities: Vector
  Fields in AdS/CFT},''
  \href{http://dx.doi.org/10.1088/1126-6708/2006/11/085}{{\em JHEP} {\bfseries
  11} (2006) 085}, \href{http://arxiv.org/abs/hep-th/0606113}{{\ttfamily
  arXiv:hep-th/0606113}}.

\bibitem{Argurio:2022vtz}
R.~Argurio and A.~Caddeo, ``{Comments on Holographic Level/Rank Dualities},''
  \href{http://arxiv.org/abs/2205.06115}{{\ttfamily arXiv:2205.06115
  [hep-th]}}.

\bibitem{Das:2022auy}
A.~Das, R.~Gregory, and N.~Iqbal, ``{Higher-form symmetries, anomalous
  magnetohydrodynamics, and holography},''
  \href{http://arxiv.org/abs/2205.03619}{{\ttfamily arXiv:2205.03619
  [hep-th]}}.

\bibitem{JADRAEGC}
J.~Aguilera~Damia, R.~Argurio, and E.~Garcia-Valdecasas, ``to appear,''.

\bibitem{Seiberg:2016gmd}
N.~Seiberg, T.~Senthil, C.~Wang, and E.~Witten, ``{A Duality Web in 2+1
  Dimensions and Condensed Matter Physics},''
  \href{http://dx.doi.org/10.1016/j.aop.2016.08.007}{{\em Annals Phys.}
  {\bfseries 374} (2016) 395--433},
  \href{http://arxiv.org/abs/1606.01989}{{\ttfamily arXiv:1606.01989
  [hep-th]}}.

\bibitem{Polyakov:1988md}
A.~M. Polyakov, ``{Fermi-Bose Transmutations Induced by Gauge Fields},''
  \href{http://dx.doi.org/10.1142/S0217732388000398}{{\em Mod. Phys. Lett. A}
  {\bfseries 3} (1988) 325}.

\bibitem{Aharony:2015mjs}
O.~Aharony, ``{Baryons, monopoles and dualities in Chern-Simons-matter
  theories},'' \href{http://dx.doi.org/10.1007/JHEP02(2016)093}{{\em JHEP}
  {\bfseries 02} (2016) 093}, \href{http://arxiv.org/abs/1512.00161}{{\ttfamily
  arXiv:1512.00161 [hep-th]}}.

\bibitem{Karch:2016sxi}
A.~Karch and D.~Tong, ``{Particle-Vortex Duality from 3d Bosonization},''
  \href{http://dx.doi.org/10.1103/PhysRevX.6.031043}{{\em Phys. Rev. X}
  {\bfseries 6} no.~3, (2016) 031043},
  \href{http://arxiv.org/abs/1606.01893}{{\ttfamily arXiv:1606.01893
  [hep-th]}}.

\bibitem{Wilczek:1981du}
F.~Wilczek, ``{Magnetic Flux, Angular Momentum, and Statistics},''
  \href{http://dx.doi.org/10.1103/PhysRevLett.48.1144}{{\em Phys. Rev. Lett.}
  {\bfseries 48} (1982) 1144--1146}.

\bibitem{Jain:1989tx}
J.~K. Jain, ``{Composite fermion approach for the fractional quantum Hall
  effect},'' \href{http://dx.doi.org/10.1103/PhysRevLett.63.199}{{\em Phys.
  Rev. Lett.} {\bfseries 63} (1989) 199--202}.

\end{thebibliography}\endgroup
\makeatother

\end{document}